\documentclass[sn-mathphys-num]{sn-jnl}

\usepackage{graphicx}%
\usepackage{multirow}%
\usepackage{amsmath,amssymb,amsfonts}%
\usepackage{amsthm}%
\usepackage{mathrsfs}%
\usepackage[title]{appendix}%
\usepackage{xcolor}%
\usepackage{textcomp}%
\usepackage{manyfoot}%
\usepackage{booktabs}%
\usepackage{algorithm}%
\usepackage{algorithmicx}%
\usepackage{algpseudocode}%
\usepackage{listings}%
\usepackage{upgreek}
\usepackage{enumerate}
\usepackage{times}
\usepackage[english]{babel}

\hypersetup{
    colorlinks,
    linkcolor={black},
    citecolor={black},
    urlcolor={black}
}

\newcommand{\beq}{\begin{equation}}
\newcommand{\eeq}{\end{equation}}
\newcommand{\bea}{\begin{eqnarray}}
\newcommand{\eea}{\end{eqnarray}}
\newcommand{\bit}{\begin{itemize}}
\newcommand{\eit}{\end{itemize}}
\newcommand{\ben}{\begin{enumerate}}
\newcommand{\een}{\end{enumerate}}
\newcommand{\nn}{\nonumber}

\def\scri{\mathscr{I}}

\newcommand{\BLcoord}{(t, r, \theta, \phi)}

\renewcommand{\d}{\,\textnormal{d}}
\newcommand{\spin}{\frak s}
\newcommand{\rh}{r_{\rm h}}
\newcommand{\rc}{r_{\rm c}}

\newcommand{\spinPsi}[1]{{}_{#1}\!\Psi}
\newcommand{\spinS}[1]{\, {}_{\spin}S_{\ell m}( #1; \theta)}

\newcommand{\spinR}[2]{\, {}_{\spin}{\cal R}_{\ell m}( #1; #2)}

\newcommand{\spinHeun}[2]{\, {}_{\spin}{\rm R}_{\ell m}( #1; #2)}
\newcommand{\spinHeunType}[3]{\, {}_{\spin}{\rm R}^{#3}_{\ell m}( #1; #2)}

\newcommand{\OutNullCoord}{(u, r, \theta, \hat \phi)}
\newcommand{\spinOutNullPsi}[1]{{}_{#1}\!\hat\Psi}
\newcommand{\spinOutNullR}[2]{\, {}_{\spin}{\hat {\cal R}}_{\ell m}( #1; #2)}

\newcommand{\InNullCoord}{(v, r, \theta, \check \phi)}

\newcommand{\spinInNullPsi}[1]{{}_{#1}\!\check\Psi}

\newcommand{\spinInNullR}[2]{\, {}_{\spin}{\check {\cal R}}_{\ell m}( #1; #2)}

\renewcommand{\t}{\tilde{t}}
\newcommand{\GenCoord}{(\t, r, \theta, \tilde \phi)}

\newcommand{\spinGenPsi}[1]{{}_{#1}\!\tilde\Psi}

\newcommand{\spinGenR}[2]{\, {}_{\spin}{\tilde {\cal R}}_{\ell m}( #1; #2)}

\renewcommand{\i}{{\rm i}}

\newcommand{\spinHyperPsi}[1]{{}_{#1}\!\bar\Psi}

\newcommand{\spinHyperR}[2]{\, {}_{\spin}{\bar {\cal R}}_{\ell m}( #1; #2)}

\theoremstyle{thmstyleone}%
\theoremstyle{thmstyletwo}%

\theoremstyle{thmstylethree}%

\begin{document}

\title{\textbf{The Confluent Heun functions in Black Hole Perturbation Theory: a spacetime interpretation}}


\author[]{ Marica Minucci\footnote{E-mail address: marica.minucci@nbi.ku.dk}\; }
\author[]{Rodrigo Panosso Macedo\footnote{E-mail address: rodrigo.macedo@nbi.ku.dk}}

\affil[]{\small \orgdiv{\textit{Niels Bohr International Academy}}, \orgname{\textit{Niels Bohr Institute}}, \orgaddress{\textit{\street{Blegdamsvej 17} \postcode{2100}, \city{Copenhagen} \country{Denmark}}}}


\abstract{
This work provides a spacetime interpretation of the confluent Heun functions within black hole perturbation theory (BHPT) and explores their relationship to the hyperboloidal framework. In BHPT, the confluent Heun functions are solutions to the radial Teukolsky equation, but they are traditionally studied without an explicit reference to the underlying spacetime geometry. Here, we show that the distinct behaviour of these functions near their singular points reflects the structure of key surfaces in black hole spacetimes. By interpreting homotopic transformations of the confluent Heun functions as changes in the spacetime foliation, we connect these solutions to different regions of the black hole’s global structure, such as the past and future event horizons, past and future null infinity, spatial infinity, and even past and future timelike infinity. We also discuss the relationship between the confluent Heun functions and the hyperboloidal formulation of the Teukolsky equation. Although neither confluent Heun form of the radial Teukolsky equation can be interpreted as hyperboloidal slices, this approach offers new insights into wave propagation and scattering from a global black hole spacetime perspective.
}




\maketitle

 {\hypersetup{linkcolor=black}
  \tableofcontents
}

\section{Introduction}
Over the past decade, the hyperboloidal framework has established itself as a fundamental approach to black hole perturbation theory (BHPT) \cite{Zenginoglu:2007jw,Zenginoglu:2011jz,jaramillo2021pseudospectrum,PanossoMacedo:2023qzp}. One of the central motivations for this formalism is the resolution of a long standing puzzle within this theory~\cite{PanossoMacedo:2024nkw}. In the post-merger dynamics, gravitational waves carry energy toward the black hole and to the infinitely far wave zone until the spacetime settles into a stationary regime. The gravitational waves signal in this ringdown phase has a very particular signature, with oscillations decaying exponentially at characteristic frequencies $\omega\in {\mathbb C}$, the so-called quasinormal modes \cite{kokkotas1999quasi,Nollert99_qnmReview,berti_quasinormal_2009,konoplya_quasinormal_2011}. The quasinormal mode problem is typically formulated as an eigenvalue problem and, associated to each frequency $\omega$, there exists a time-harmonic quasinormal mode function $\Psi$ defined on an underlying Riemannian manifold with topology ${\mathbb R}\times \mathcal{S}^2$. In the traditional mathematical approach, these quasinormal mode eigenfunctions $\Psi$ grow exponentially towards the black hole and infinity. It is therefore puzzling that black hole stability implies time decay for linearised perturbations, but the associated time-harmonic perturbations is singular in the asymptotic regions.

A resolution to this conundrum arises when the global structure of the spacetime is fully taken into account. It becomes clear that the singular behaviour of the quasinormal mode eigenfunctions is a mere consequence of the choice of coordinates employed, in particular the notion of constant time hypersurface yielding the Riemannian manifold. The typical choice for a time coordinate $t$ is actually singular at the event horizon and at infinity. In other words, surfaces of $t=$ constant accumulate at the bifurcation sphere $\mathcal{B}$ and at spatial infinity ${i}^0$.  Reformulating the problem in terms of regular spacelike surfaces extending between the black-hole event horizon ${\cal H}^+$ and null infinity ${\scri}^+$ yields regular quasinormal mode eigenfunctions, as intuitively expected.

The hyperboloidal framework has led to great advances in the BHPT, see e.g. \cite{PanossoMacedo:2023qzp,PanossoMacedo:2024nkw} for reviews. All applications focus on scenarios where energy flows {\em exclusively} across ${\cal H}^+$ and ${\scri}^+$, with this property being the core feature defining the asymptotic behaviour of the field. However, there are further scenarios of interest in BHPT for which the field's behaviour at other geometrical surfaces --- such as ${\cal H}^-$ and ${\scri}^-$ --- play an important role. 

As a first example, let us consider the traditional definition of quasinormal modes as the poles of the Green's function associated with the underlying wave operator~\cite{kokkotas1999quasi,Nollert99_qnmReview}. Broadly speaking, to construct the correct Green's function at a non-quasinormal mode frequency, one needs two linearly independent solutions to the homogenous wave equation, with each individual field satisfying the desired boundary conditions. In BHPT, these solutions are referred to as the ``in" and ``up" solutions, and they contain the field's behaviour either at $\scri^-$ or ${\cal H}^-$, see e.g.~\cite{Chandrasekhar83,Pound:2021qin}. For $\omega\in {\mathbb R}$, the ``in" solution describes scattering of monochromatic waves by the black hole, and the corresponding transmission and reflexion coefficients provides valuable information about the spacetime. 

Currently, the hyperboloidal framework in BHPT still lacks a comprehensive formalism and numerical infrastructure to tackle problems where these other asymptotic regions are also relevant. Developing this infrastructure is necessary to computationally address open problems in the field such as the relation between the traditional definition of quasinormal modes as the poles of the Green's~\cite{kokkotas1999quasi,Nollert99_qnmReview} function and more recent approaches that interpret quasinormal modes as a formal eigenvalue problem on an appropriate Hilbert space~\cite{Gajic:2019oem,Gajic:2019qdd,Gajic:2024xrn}; understanding notions of quasinormal modes orthogonality \cite{Zimmerman:2014aha,Green:2022htq,London:2023aeo} from the hyperboloidal perspective; or taking the fully global structure of the spacetime into account when studying wave scattering in black hole spacetimes. 

This work presents initial steps towards filling this gap. For this purpose, we consider the Teukolsky equation in the Kerr spacetime, and we review the representation of solutions to the radial equation in terms of the confluent Heun function~\cite{Ronveaux95}. 

A decade after Teukolsky's seminal works \cite{teukolsky1972rotating,teukolsky1973perturbations,Teukolsky:1974yv}, it was pointed out \cite{Marcilhacy1983} that the underlying radial and angular equations are, in fact, particular forms of a single linear differential equation resulting from the coalescence of singular points. Moreover, these equations may be obtained by elementary transformations from the confluent Heun function~\cite{Blandin1983}. Similar techniques were also employed when discussing  the analytic representation for the Teukolsky solutions~\cite{Leaver:1986vnb}. In the past decade, there has been an increase of interest in the Heun functions in the context of BHPT, see \cite{Hortacsu:2011rr} and references within for a review. Not only these functions provide an analytical framework for studying key observables in BHPT, such as quasinormal modes frequencies, Hawking radiation, and greybody factor (non exhaustive e.g. \cite{Vieira:2016ubt,Noda:2022zgk}), but they also relate to novel techniques in BHPT to understand quasinormal modes' properties such as spatial completeness and orthogonality~\cite{London:2020uva,London:2023aeo, London:2023idh}.

In this work, we offer a spacetime interpretation for the Heun's confluent solution in BHPT by considering its relation with the underlying coordinates system parametrising the spacetime. In a nutshell, the standard theoretical approach to the \textit{confluent Heun equation} (CHE) always focuses on the linear ordinary differential equation (ODE)
 \bea
 \label{eq:canonoical_CHE_intro}
 \dfrac{d^2 Z(z)}{dz^2} + \left(  \dfrac{\gamma}{z} + \dfrac{\delta}{z-1} + \epsilon \right)  \dfrac{d Z(z)}{dz} +  \dfrac{\alpha z - q}{z(z-1)}  Z(z) =0,
 \eea
also known as the CHE in the canonical form. The equation has regular singular points at $z = 0$ and $z =1$, and linearly independent local solutions around any of these points have the leading order behaviour  
\bea
Z(z) \sim 1 + {\cal O}\big(z\big),  &\quad& Z(z) \sim z^{1-\gamma}\left( 1 + {\cal O}\big(z\big)\right), \label{eq:z0_behaviour} \\
Z(z) \sim 1 + {\cal O}\big(z-1\big),  &\quad& Z(z) \sim (z-1)^{1-\delta}(1 + {\cal O}\big(z-1\big)). \label{eq:z1_behaviour}
\eea
The equation has also an irregular singular point at $z\rightarrow \infty$, with linearly independent solutions around it resulting from an asymptotic expansion with leading order behaviour
\bea
\label{eq:asymp_beh}
Z(z) \sim \dfrac{1}{z^{\alpha/\epsilon}}\bigg( 1 + {\cal O}\big(1/z\big)\bigg) , \quad Z(z) \sim \dfrac{e^{-\epsilon z}}{z^{\gamma+\delta - \alpha/\epsilon}}\bigg( 1 + {\cal O}\big(1/z\big)\bigg).
\eea
The full solution to eq.~\eqref{eq:canonoical_CHE_intro} is then represented as a linear combination of the two independent local solutions around the regular singular points as given by expressions \eqref{eq:z0_behaviour} or \eqref{eq:z1_behaviour}. Alternatively, the full solution can also be represented as a linear combination of the independent asymptotic expansions \eqref{eq:asymp_beh}.

Further representations of the canonical form \eqref{eq:canonoical_CHE_intro} arise via the so-called $s$-homotopic transformation, which provides a continuous mapping between functions via 
\beq
\label{eq:homo_intro}
Z(z) = e^{\nu z} z^{\mu_0} (z-1)^{\mu_1} \bar Z(z).
\eeq
A useful qualitative description of the effects from transformation \eqref{eq:homo_intro} is to reduce the exponents of expansions \eqref{eq:z0_behaviour} and \eqref{eq:z1_behaviour} by $\mu_0$ and $\mu_1$, respectively. The asymptotic behaviour \eqref{eq:asymp_beh} is also reduced by a factor $e^{\nu z} z^{\mu_0 + \mu_1}$. 

While such approach has its advantages at an analytical level, it misses the geometrical interpretation from the spacetime perspective. Since in BHPT the CHE actually results from projecting a wave equation into the frequency domain via a Fourier (or Laplace) transform\footnote{Frequency domain equations also result from a two-time scale expansion of Einstein's field equation \cite{Miller:2020bft}.}, the representation  \eqref{eq:canonoical_CHE_intro} implicitly assumes an underlying notion of coordinate time within the coordinate system parameterising the Lorentzian manifold. The homotopic transformation \eqref{eq:homo_intro} is, therefore, a consequence of changing the spacetime foliation. As a result, the changes in the behaviour around the singular points $z=0$, $z=1$ and $z\rightarrow \infty$ reflects the fact that these surfaces are actually different geometrical loci in the global structure of the spacetime --- see sec.~\ref{sec:CHE_Spacetime}.

In the next section we discuss these properties in more detail. Sec.~\ref{sec:CHE} reviews some basic properties of the CHE, whereas sec.~\ref{sec:Kerr} reviews the Kerr spacetime and the common usage of the CHE in BHPT. We provide the spacetime interpretation for the confluent Heun solutions in sec.~\ref{sec:CHE_Spacetime}, whereas sec.~\ref{sec:hyp} compares these interpretations against hyperboloidal foliations before our conclusion in sec.~\ref{sec:conclusion}.

\subsection{Notation}
We employ lower case latin letters to represent tensorial quantities in the abstract index notation, whereas Greek letters indicate the components of the tensor in a given coordinate system. Thus, $\ell^a$ and $\hat \ell ^a$ are different vectors, whereas $\ell^\mu$ and $\ell^{\hat \mu}$ are the components of the same vector $\ell^a$, but expressed in the coordinate systems $x^\mu$ or $\hat x^\mu$, respectively. 

We denote the null tetrad basis via $(\ell_+^a, \ell_-^a, m^a, m^\star{}^a)$, with $\ell_+^a$ and $\ell_-^a$ associated, respectively, with outgoing and ingoing null directions in a black hole, asymptotically flat spacetime. We avoid the typical Newman-Penrose notation for the null tetrad basis because the signs $\ell_\pm$ will be associated with particular choices for the representation of the radial Teukolsky equation as a confluent Heun function. Besides, the vector $n^a$ is reserved for the normal vector to a given hypersurface of constant time coordinate. 

The imaginary unit is denoted by $\i$, whereas the symbol $i=0,1$ is reserved for enumerating the singular points of the Heun Equation. We employ geometric units with $G=c=1$.


\section{The Confluent Heun Equation}\label{sec:CHE}
Following ref.~\cite{Ronveaux95}, we begin with an introductory discussion about the confluent Heun equation and its solutions. The literature and results on this topic are vast, and a detailed analysis goes beyond the scope of this work (see  \cite{Ronveaux95} for a complete review). Apart from reviewing some relevant properties and results, this section also serves the purpose of fixing the notation employed in this work.

The Heun equation is a second-order linear differential equation with four regular singular points, with the confluent Heun equation (CHE) arising when two singularities are brought into coincidence. The general form of the CHE reads
\beq
\label{eq:GenConfHeun}
\dfrac{d^2}{dy^2} Y(y) + \left( \sum_{i=0}^1 \dfrac{A_i}{y-y_i} + E \right) \dfrac{d}{dy} Y(y) + \left( \sum_{i=0}^1 \dfrac{C_i}{y-y_i}  + \sum_{i=0}^1 \dfrac{B_i}{(y-y_i)^2} + D \right) Y(y) = 0,
\eeq
with $y\in[0,\infty)$. The regular singular points $y_0$ and $y_1$ are arbitrary, and so are the coefficients $A_i, B_i, C_i$ ($i=0,1$), $D, E$, whereas $y\rightarrow \infty$ is an irregular singular point.

The two linear independent solutions to eq.~\eqref{eq:GenConfHeun} are represented either locally around any regular singular point $y_i$, or as an asymptotic expansion around $y\rightarrow \infty$ via
\beq
\label{eq:SolBehav_Y}
Y(y) \sim \left\{ 
\begin{array}{ccc}
{\cal K}_{i+} (y - y_i)^{\varrho^+_i} + {\cal K}_{i-} (y - y_i)^{\varrho^-_i}, & {\rm for} & y\rightarrow y_i \\
{\cal K}_{\infty+} \,  \dfrac{ e^{  y \varsigma^+} }{y^{\eta^+}} + {\cal K}_{\infty-}\,  \dfrac{e^{  y \varsigma^-}}{ y^{\eta^-}}  & {\rm for} & y\rightarrow \infty \\
\end{array}
\right.
\eeq
with the ${\cal K}_\pm$'s constants, eventually determined by the boundary conditions, and the characteristic exponents $\varrho^\pm_i$, $\varsigma^\pm$ and $\eta^\pm $ are given by
\bea
\label{eq:charc_exp_gen_rho}
\varrho^\pm_i &=& \dfrac{1-A_i}{2} \pm \sqrt{ \left(  \dfrac{ 1 - A_i  }{2}   \right)^2  - B_i     }, \\
\label{eq:charc_exp_gen_sig}
\varsigma^\pm &=& -\dfrac{E}{2} \pm \dfrac{1}{2}\sqrt{E^2 - 4 D}, \\
\label{eq:charc_exp_gen_eta}
\eta^\pm &=& \sum_{i=0}^1 \left( \dfrac{A_i}{2} \pm \dfrac{C_i - E A_i /2}{\sqrt{E^2 - 4 D}} \right).
\eea 
Despite the generic form of eq.~\eqref{eq:GenConfHeun}, the CHE are usually expressed in the literature in different forms, with a more standard representation in terms of the so-called \emph{non-symmetrical canonical form}. Alternatively, one may also map the independent variable $y\rightarrow 1/y$ to explicitly incorporate the irregular singular point into the equation. We discuss both approaches in the next sections.

\subsection{Non-symmetrical canonical form}\label{sec:HeunCanonical}
 The non-symmetrical canonical form of the CHE \eqref{eq:canonoical_CHE_intro} arises from eq.~\eqref{eq:GenConfHeun} when the generic regular singular points $y_0$ and $y_1$ are mapped into $z_0 = 0$ and $z_1 = 1$, respectively, via
 %
 \beq
 \label{eq:change_coord_y_z}
 z = \dfrac{y - y_0}{y_1 - y_0} \Longleftrightarrow y = y_0 + (y_1 - y_0)z,
 \eeq 
together with an s-homotopic transformation 
\beq
 \label{eq:homo_y_z}
Y(y(z)) =e^{\nu z} z^{\mu_0} (z-1)^{\mu_1} Z(z).
\eeq
Eqs.~\eqref{eq:change_coord_y_z} and \eqref{eq:homo_y_z} combined lead to a transformation of the coefficients
\bea
\label{eq:coef_trasfo}
& A^{(z)}_i = A^{(y)}_i + 2 \mu_i, \quad B^{(z)}_i = B^{(y)}_i  + \mu_i \left(A^{(y)}_i + \mu_i - 1\right),    \\ 
& C^{(z)}_i = \left(C^{(y)}_i + \mu_i E^{(y)} \right)\left(y_1 - y_0\right) + \nu \left(  A^{(y)}_i + 2 \mu_i\right) + (-1)^i  \left( \mu_i A_j + \mu_j A_i \right) + (-1)^{i+1} 2 \mu_i \mu_j,  \nn\\
& D^{(z)} = D^{(y)}(y_1 - y_0)^2  + \nu E^{(y)}   (y_1 - y_0) + \nu^2, \quad
E^{(z)} = E^{(y)} \left( y_1 - y_0\right) + 2 \nu,  \nn
\eea
with $i,j= 0,1$ and $i\neq j$, while the superscripts $^{(y)}$ and $^{(z)}$ denote the underlying independent variable the coefficients relate to.

The non-symmetrical canonical form \eqref{eq:canonoical_CHE_intro} results from imposing $B^{(z)}_i =D^{(z)} = 0$ in \eqref{eq:coef_trasfo}, achieved via the homotopic transformation \eqref{eq:homo_y_z} with the parameters
\beq
\label{eq:homo_coeff_fix}
\mu_i = \dfrac{1- A^{(y)}_i}{2} \pm \sqrt{  \left( \dfrac{1-A^{(y)}_i}{2}  \right)^2 - B^{(y)}_i}, \quad \nu =\dfrac{y_1-y_0}{2} \left(  -E^{(y)} \pm \sqrt{  E^{(y)}{}^2 - 4 D^{(y)}  }  \right).
\eeq
 The canonical parameters are then given by
 \beq
 \label{eq:canonical_par}
 \gamma = A^{(z)}_0, \quad \delta  = A^{(z)}_1, \quad \epsilon = E^{(z)}, \quad \alpha = \dfrac{C^{(z)}_0 + C^{(z)}_1}{E^{(z)}}, \quad q = C^{(z)}_0.
 \eeq
The function $Z(z)$'s behaviour around the singular points and asymptotically is equivalent to eq.~\eqref{eq:SolBehav_Y} with characteristic exponents
\bea
\label{eq:CharExp_rho}
&\varrho_0^+ = 1-\gamma, \quad \varrho_1^+ = 1-\delta, \quad \varrho_0^- = \varrho_1^- = 0, \\ 
\label{eq:CharExp_sigeta}
& \varsigma^+ = 0, \quad  \varsigma^- = -\epsilon,  \quad \eta^+ = \dfrac{\alpha}{\epsilon}, \quad \eta^- = \gamma + \delta - \dfrac{\alpha}{\epsilon}.
\eea

We end this section emphasising the important role played by the homotopic transformation~\eqref{eq:homo_y_z} when expressing the CHE in its canonical form \eqref{eq:canonoical_CHE_intro}. In fact, the canonical form is not unique, since eq.~\eqref{eq:homo_coeff_fix} provides a total of 8 different possible transformations of the generic CHE \eqref{eq:GenConfHeun} into the canonical CHE \eqref{eq:canonoical_CHE_intro}. 

Besides, there also exist 6 transformations of the independent variable $z$ mapping 2 of the 3 points $0, 1, \infty$ into $0,1$. We comment on that approach in the next section. 

\subsection{An alternative form of the Confluent Heun Equation}\label{sec:NaturalHeun}
The confluent Heun equation \eqref{eq:GenConfHeun} admits an alternative general form, obtained by
introducing the transformation
\bea
\label{eq:compact_y_sig}
\sigma = 1/y
\eea
mapping the original singular points $y_i$ into $\sigma_i = 1/y_i$ ($i=0,1$), and the irregular singular point $y\rightarrow \infty$ into $\sigma_2 = 0$. Besides, the ordinary point $y=0$ is pushed to $\sigma \rightarrow \infty$. 

From eq.~\eqref{eq:compact_y_sig}, we obtain the following alternative general form of the CHE for $\Upsilon(\sigma) = Y(y(\sigma))$
 \bea
 \label{eq:CHE_compact}
 \dfrac{d^2 {\Upsilon}(\sigma)}{d\sigma^2} + \bigg{(} \sum_{i=0}^{1} \dfrac{A{}_i{}^{(\sigma)}}{\sigma- \sigma_i} &+& \sum_{j=1}^{2}  \dfrac{E{}_{j}{}^{(\sigma)} }{(\sigma - \sigma_2)^j}\bigg{)}  \dfrac{d \Upsilon(\sigma)}{d\sigma} + \nn \\
 &+& \Bigg{(} \sum_{i=0}^{1}\left( \dfrac{C{}_i{}^{(\sigma)}}{\sigma - \sigma_i}+ \dfrac{B{}_i{}^{(\sigma)}}{(\sigma - \sigma_i)^2} \right)+\sum_{j=1}^{4}  \dfrac{D{}_{j}{}^{(\sigma)} }{(\sigma - \sigma_2)^j} \Bigg{)}\Upsilon(\sigma) =0.
 \eea 
For the point at infinity to be an ordinary point of eq. (\ref{eq:CHE_compact}), the coefficients must satisfy the following conditions
\bea
\label{eq:cond_Allternative_Heun}
&\sum\limits_{i=0}^{1} A{}_i{}^{(\sigma)} + E{}_{1}{}^{(\sigma)}=2, \qquad \sum\limits_{i=0}^{1}C{}_i{}^{(\sigma)}+ D{}_{1}{}^{(\sigma)}=0, \nn \\
&\sum\limits_{i=0}^{1}B{}_i{}^{(\sigma)}+D{}_{2}{}^{(\sigma)}+\sum\limits_{i=0}^{1}C{}_i{}^{(\sigma)}\sigma_i + \sigma_2 D{}_{1}{}^{(\sigma)}=0,\\
& \sum\limits_{i=0}^{1} 2B{}_i{}^{(\sigma)}\sigma_i +\sum\limits_{i=0}^{1}C{}_i{}^{(\sigma)} \sigma{}_i^2 + \sigma{}_2^2 D{}_{1}{}^{(\sigma)}+ 2\sigma{}_2 D{}_{2}{}^{(\sigma)}+ D{}_{3}{}^{(\sigma)}=0. \nn
\eea
These conditions reduce the number of independent parameters in equation (\ref{eq:CHE_compact}) to $11$, including those which determine the location of the singular points. 

As a side remark, an equation with the same form as eq.~\eqref{eq:CHE_compact} arises if the compactification is performed in terms of $\sigma = 1/z$, cf.~\eqref{eq:change_coord_y_z}. Though the resulting coefficients $A{}_i{}^{(\sigma)}$, $B{}_i{}^{(\sigma)}$, $C{}_i{}^{(\sigma)}$, $D{}_j{}^{(\sigma)}$ and $E{}_j{}^{(\sigma)}$ would assume different values, the conditions \eqref{eq:CharExp_rho} would still hold. Both compactifications  are employed when considering the Teukolsky equation in the hyperboloidal minimal gauge, via different choices of the radial transformation, cf. sec.~\ref{sec:hyp}. In the next section, we review how the CHE appears in black hole perturbation theory, and explicitly show how to understand the CHE transformations in terms of different hypersurfaces of constant time in the black hole spacetime.

\section{Kerr spacetime}\label{sec:Kerr}
In this section, we review the common approach to the usage of the confluent Heun equation in the context of black hole perturbation theory. The background spacetime is given by the Kerr solution, and the perturbation dynamics are dictated by the Teukolsky equation \cite{teukolsky1973perturbations}.

\subsection{Boyer-Lindquist Coordinates}
In black hole perturbation theory, the line element for the Kerr spacetime is traditionally expressed in terms of the Boyer-Lindquist (BL) coordinates $x^\mu = (t,r,\theta,\phi)$,
\bea
\label{eq:Kerr_BL}
\d s^2 =& -f(r, \theta) \d t^2 - \dfrac{4Mar}{\Sigma(r, \theta)}\sin^2\theta \d t \d\varphi + \dfrac{\Sigma(r, \theta)}{\Delta(r)} \d r^2 \\
&+ \Sigma (r, \theta)\d\theta^2   + \sin^2\theta\left( \Sigma_0 (r)+ \dfrac{2Ma^2r}{\Sigma(r, \theta)}\sin^2\theta \right)\d\phi^2 \nn
\eea
with
\bea
&&\Delta(r) =  r^2 -2Mr + a^2 = (r-\rh )(r-\rc), \label{eq:Delta}  \\
\label{eq:Sigmas}
&&\Sigma(r,\theta) = r^2 + a^2\cos^2\theta, \quad  \Sigma_0(r) = \Sigma(r,0) = r^2 + a^2, \\
&&f(r,\theta) = 1 - \dfrac{2Mr}{\Sigma(r,\theta)}, \quad f_0(r) = f(r,0) \label{eq:f}.
\eea
The parameters $M$ and $a$ relate, respectively, to the black hole's mass and angular momentum. The condition $\Delta(r) = 0$ defines the event ($\rh$) and Cauchy ($\rc$) horizons
\bea
\dfrac{\rh}{M} =  1 + \sqrt{1 - \dfrac{a^2}{M^2}}, \quad \dfrac{\rc}{M} =  1 - \sqrt{1 - \dfrac{a^2}{M^2}}.
\eea

Some calculations are more convenient performed with dimensional quantities measured in terms of the event horizon length scale $r_h$, as opposed to the more common choice using the mass parameter $M$. For that purpose, we introduce a dimensionless black hole spin parameter $\kappa$ equivalently defined by
\beq
\kappa = \dfrac{a}{r_h}=\dfrac{a}{M + \sqrt{M^2 - a^2}}, \quad \kappa^2 = \dfrac{\rc}{\rh}.
\eeq
With this definition, the mass parameter reads $2M = \rh \left(1+\kappa^2\right)$.

Further important quantities for the Kerr metric are the horizons angular velocities 
\bea
\Omega_{\rm h} = \dfrac{a}{2M \rh} = \dfrac{\kappa}{\rh (1+\kappa^2)}, \quad \Omega_{\rm c} = \dfrac{a}{2M \rc} = \dfrac{1}{\rh \kappa (1+\kappa^2)},
\eea
and the horizons surface gravities
\beq
\varkappa_{\rm h} = \dfrac{1}{4M} - M \Omega_{\rm h} ^2 = \dfrac{1-\kappa^2}{2\rh (1+\kappa^2)} , \quad \varkappa_{\rm c} = \dfrac{1}{4M} - M \Omega_{\rm c} ^2 = - \dfrac{1-\kappa^2}{2\rh \kappa^2 (1+\kappa^2)}.
 \eeq
In terms of these observables, the tortoise coordinate $r^*(r)$ reads
\beq
\label{eq:tort_r}
\dfrac{dr^*(r)}{dr}=\dfrac{r^2 +a^a}{\Delta(r)} \quad \Rightarrow \quad  r^*(r) = r + 2 M \ln\left( \dfrac{r}{\rh} \right) + \dfrac{1}{2 \varkappa_{\rm h}} \ln\left(1-  \dfrac{\rh}{r} \right)  + \dfrac{1}{2 \varkappa_{\rm c}} \ln\left(1 -  \dfrac{\rc}{r} \right),
\eeq
whereas the ``tortoise" azimutal angular phase $\chi(r)$ reads
 \beq
 \label{eq:tort_chi}
 \dfrac{d\chi(r)}{dr} = \dfrac{a}{\Delta(r)} \quad \Rightarrow \quad \chi(r) = \dfrac{\Omega_{\rm h}}{2\varkappa_{\rm h}} \ln\left(1-  \dfrac{\rh}{r} \right) + \dfrac{\Omega_{\rm c}}{2\varkappa_{\rm c}} \ln\left(1-  \dfrac{\rc}{r} \right)
  \eeq

\subsection{The Teukolsky equation}
The equation governing the dynamics of perturbations around this spacetime is given by the (sourceless) Teukolsky equation. When expressed in terms of the Boyer-Lindquist coordinates $(t, r, \theta, \phi)$ it reads
\bea
& 0= - \left[  \dfrac{(\Sigma_0(r))^2}{\Delta(r)} - a^2 \sin^2\theta \right]  \spinPsi{\spin}_{,tt} (x^a)- \dfrac{4Mar}{\Delta(r)}  \spinPsi{\spin}_{,t\phi}(x^a) - \left[ \dfrac{a^2}{\Delta(r)} - \dfrac{1}{\sin^2\theta} \right] \spinPsi{\spin}_{,\phi\phi}  (x^a)\nn \\
\label{eq:Teukolsky_BL}
& + \Delta^{-\spin}(r) \, \partial_r \bigg( \Delta^{\spin+1}(r) \,  \spinPsi{\spin}_{,r}(x^a) \bigg)  + 2\spin\left[ \dfrac{M(r^2-a^2)}{\Delta(r)} + (r+\i a\cos\theta) \right] \spinPsi{\spin}_{,t}(x^a)  \label{eq:Teukolsky_BL}   \\
&+2 \spin \left[ \dfrac{a(r-M)}{\Delta(r)}  + \i\dfrac{\cos\theta}{\sin^2\theta}\right] \spinPsi{\spin}_{,\phi}(x^a) + \dfrac{1}{\sin\theta}\partial_\theta \bigg( \sin\theta \,  \spinPsi{\spin}_{,\theta}(x^a)\bigg) - \spin(\spin\cot^2\theta -1)\spinPsi{\spin}(x^a). \nn
\eea
This equation is formulated for the master function $\spinPsi{\spin}$, where $\spin$ denotes the field's spin-weight parameter. The case $\spin=0$ describes a scalar perturbation, whereas $\spin=\pm 1$ and $\spin=\pm 2$ describe electromagnetic and gravitational perturbations, respectively. 

In this work, we will focus on the gravitational degrees of freedom, where $\spin=\pm 2$ is related to the Newman-Penrose components of the Weyl curvature tensor $\Psi_0$ and $\Psi_4$ via 
\beq
\label{eq:TeukMaster_NewmanPenrose}
\spinPsi{+2}\BLcoord = \Psi_0\BLcoord, \quad \spinPsi{-2}\BLcoord = \left( r + \i a \cos\theta \right)^{4} \Psi_4\BLcoord.
\eeq
The Newman-Penrose curvature components $\Psi_0$ and $\Psi_4$ are defined in terms of the \textit{Kinnersley null tetrad} $(\ell^a_+, \ell^a_-, m^a, m^\star{}^a)$, whose components in the BL coordinates $x^\mu = (t,r,\theta,\phi)$ read
\bea
\label{eq:Kin_null_tetrad_lk}
&\ell^\mu_+ = \dfrac{r^2+a^2}{\Delta(r)}\delta^\mu_t + \delta^\mu_r + \dfrac{a}{\Delta(r)}\delta^\mu_\phi, \quad \ell_-^\mu =\dfrac{1}{2\Sigma(r, \theta)}\bigg( \left(r^2+a^2\right)\delta^\mu_t -\Delta(r) \, \delta^\mu_r + a\,\delta^\mu_\phi\bigg), \\
\label{eq:Kin_null_tetrad_m}
& m^\mu = \dfrac{1}{{\sqrt{2} (r+ \i a \cos\theta)}} \left( \i a \sin\theta \delta^\mu_t + \delta^\mu_\theta + \dfrac{\i}{\sin\theta}\delta^\mu_\phi \right).
\eea
One can separate the Teukolsky equation in the frequency domain by introducing a Fourier decomposition
\beq
\label{eq:Fourier_BL}
\spinPsi{\spin}(t, r, \theta, \phi) = \dfrac{1}{2 \pi} \int_{-\infty }^{\infty } \sum_{\ell = |\spin|}^{\infty}\sum_{m=-\ell}^{\ell} e^{-\i \omega t} e^{\i m \phi} \spinR{\omega} {r}\spinS{\omega} {\rm d} \omega.
\eeq
This leads to two second-order ordinary differential equations for the unknown functions $\spinR{\omega}{r}$ and $\spinS{\omega}$, namely the \textit{angular Teukolsky equation}
\bea
 &\dfrac{1}{\sin\theta} \dfrac{\partial}{\partial \theta}\left( \sin\theta \dfrac{\partial}{\partial \theta} \right) \spinS{\omega} + \nn \\ 
 &+\bigg{(}a^2\omega^2\cos^2\theta - 2\spin a\omega\cos\theta-\left(\dfrac{m+\spin\cos\theta}{\sin\theta}\right)^2+\spin+A_{\ell m}\bigg{)}\spinS{\omega}=0, \label{eq:SpinWeightSphHarm}
\eea
and the \textit{radial Teukolsky equation}
\bea
\label{eq:RadialTeuk_BL} 
 &\Delta^{-\spin}(r) \, \dfrac{\d}{\d r} \left( \Delta^{\spin+1}(r) \, \dfrac{\d }{\d r}  \right) \spinR{\omega}{r} + \left( 2i\spin\omega r -a^2\omega^2 -A_{\ell m} \right)\spinR{\omega}{r}\\
&  +   \dfrac{ (\omega\Sigma_0 )^2 - 4Mam\omega r + a^2m^2+ 2i\spin\left[ am(r- M)-M\omega(r^2-a^2) \right] }{\Delta(r)}\spinR{\omega}{r}=0, \nn
\eea
where $A_{\ell m}$ are separation constants.
Both equations have two regular singular points at a finite value of the independent variable and one irregular singularity at infinity. In particular, Equation ($\ref{eq:SpinWeightSphHarm}$) is a Sturm-Liouville eigenvalue problem, and the solutions ${}_\spin S_{lm} (\theta)$ are the so-called \textit{spin-weighted spheroidal harmonics} \cite{teukolsky1973perturbations}. Here, we will focus on the radial Teukolsky equation.
\begin{figure*}[t] \centering
\includegraphics{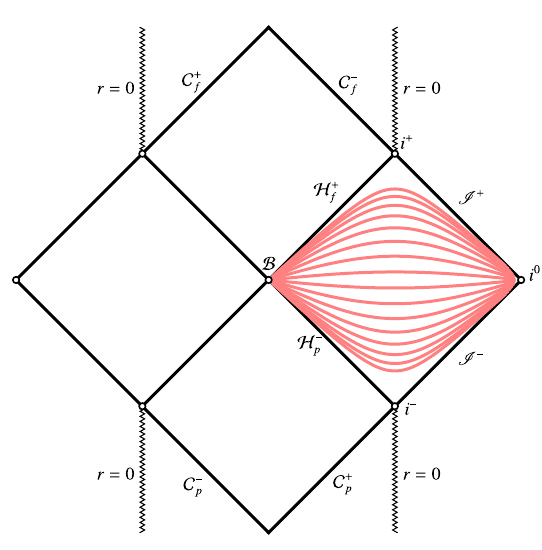}
  \caption{Carter-Penrose diagram representing hypersurfaces of constant Boyer-Lindquist time coordinate, upon which the radial Teukolsky function $\spinR{\omega} {r}$ is defined. The corresponding radial Teukolsky equation assumes the generic form of the confluent Heun equation \eqref{eq:GenConfHeun}. The derived characteristic exponent \eqref{eq:BL_coeff_C}-\eqref{eq:BL_coeff_infty} relate to the function's asymptotic behaviour as $r\rightarrow \rh$ (or $\rc$) towards the bifurcation sphere ${\cal B}$, and $r\rightarrow \infty$ at spatial infinity $i^0$.  }
    \label{fig:penrose_diagram_radial_fixing}
\end{figure*}

\smallskip
After dividing eq.~\eqref{eq:RadialTeuk_BL} by $\Delta(r)$, the radial Teukolsky equation acquires the same form as the generic CHE \eqref{eq:GenConfHeun} with coefficients given by
\bea
\label{eq:coeff_ADE}
&& A_i = 1 + \spin, \quad D= \omega^2, \quad E = 0, \\
\label{eq:coeff_B}
&& B_i = m^2\dfrac{ \kappa^2}{(1-\kappa^2)^2} - 2\rh \omega \,m \,\kappa^{3-2i}  \dfrac{ (1+\kappa^2)}{(1-\kappa^2)^2} +  \left( \rh \omega \kappa^{2-2i} \dfrac{1+\kappa^2}{1-\kappa^2} \right)^2 \nn \\
&& +(-1)^{1+i}  \, \dfrac{\i\,\spin}{1-\kappa^2} \bigg( m \kappa - \rh\omega \kappa^{2(1-i)} (1+\kappa^2)\bigg), \\
\label{eq:coeff_C}
&& C_i = \dfrac{(-1)^i}{\rh \left(1-\kappa^2\right)} \Bigg[A_{\ell m} - 2 \i \spin \kappa^{2-2i} \rh \omega +  \dfrac{  2 m^2 \kappa^2 +  2m \kappa \, \rh \omega (1+\kappa^2)^2}{(1-\kappa^2)^2}             \nn \\
&& + \bigg(  -(1+i)    -  (4 - 9i)   \kappa^2   - (5 - 9i) \kappa^4 + (2 - i) \kappa^6   \bigg) \dfrac{\kappa^{2(1-i)}\, \rh^2 \omega^2 }{(1-\kappa)^2}   \Bigg],
\eea
with the indexes $i=0,1$ enumerating the singular points $r_0 = \rc$ and $r_1 = \rh$. From the above coefficients, eqs.~\eqref{eq:charc_exp_gen_rho}-\eqref{eq:charc_exp_gen_eta} provide the characteristic exponents
\bea
\label{eq:BL_coeff_C}
&&\varrho_{{\cal C}_{\rm f}} = -\dfrac{\i \omega}{ 2\varkappa_c} + \dfrac{\i m \Omega_c}{ 2\varkappa_c} - \spin, \quad \varrho_{{\cal C}_{\rm p}} = \dfrac{\i \omega}{ 2\varkappa_c} - \dfrac{\i m \Omega_c}{ 2\varkappa_c}, \\
\label{eq:BL_coeff_H}
&&\varrho_{{\cal H}_{\rm p}} = \dfrac{\i \omega}{ 2\varkappa_h} - \dfrac{\i m \Omega_h}{ 2\varkappa_h}, \quad \varrho_{{\cal H}_{\rm f}} = -\dfrac{\i \omega}{ 2\varkappa_h} + \dfrac{\i m \Omega_h}{ 2\varkappa_h} -\spin, \\
\label{eq:BL_coeff_infty}
&& \varsigma^\pm = \pm \i \omega,  \quad \eta^+ = 1 - 2M  \omega \, \i + 2\spin,  \quad \eta^- = 1 + 2  M   \omega\, \i
\eea 
so that
\bea
\label{eq:R_BL_asymp}
\spinR{\omega}{r} = \left\{ 
\begin{array}{cccc}
{\cal K}_{{\cal H}_{\rm p}} \,(r-\rh)^{\varrho_{ {\cal H}_{\rm p } }  } &+& {\cal K}_{{\cal H}_{\rm f}} \, (r-\rh)^{ \varrho_{ {\cal H}_{\rm f } } }& (r\rightarrow \rh) \\
\\
 {\cal K}_{ \infty^+} \, r^{-\eta^+} e^{\varsigma^+ r} &+  &  {\cal K}_{{ \infty}^-} \, r^{-\eta^-} e^{\varsigma^- r} & (r\rightarrow \infty) 
\end{array}
\right.
\eea
The hypersurfaces of $t= {\rm constant}$, upon which $\spinR{\omega} {r}$ is defined, are shown in Figure~\ref{fig:penrose_diagram_radial_fixing}.

The common choices for linearly independent homogenous solutions to the radial Teukolsky equation are the ``in", ``out", ``up" and ``down". Each of them are characterised by fixing to zero one of the asymptotic behaviours, i.e.
\bea
\label{eq:R_in_and_out_BL}
{\rm (in)} \, {\cal K}_{{\cal H}_{\rm p}} = 0, \quad {\rm (out)} \, {\cal K}_{{\cal H}_{\rm f}} = 0, \quad {\rm (up)} \, {\cal K}_{ \infty^-} = 0, \quad {\rm (down)} \, {\cal K}_{ \infty^+} = 0.
\eea

Even though the radial Teukolsky equation is already represented by a CHE, from which one can derive the well-known asymptotic behaviour for the radial solution, eq.~\eqref{eq:RadialTeuk_BL} is still written in the generic form \eqref{eq:GenConfHeun}, i.e. with $B_i$ as in eqs.~\eqref{eq:coeff_B}. However, studies of the Teukolsky equation in terms of Heun function are usually presented in the literature in the {\em canonical confluent Heun equation form} \eqref{eq:canonoical_CHE_intro}. 

The canonical CHE follows from a transformation akin to eqs.~\eqref{eq:change_coord_y_z} and \eqref{eq:homo_y_z}, i.e. a coordinate change
\beq
\label{eq:z_of_r_BH}
z = \dfrac{r-\rc}{\rh - \rc},
\eeq
mapping $r = \{\rc, \rh, \infty\}$ into $z=\{0,1, \infty \}$, together with an s-homotopic transformation  
\beq
\label{HomTrasfR}
\spinR{\omega}{r(z)} =z^{\mu_c}  (z - 1)^{\mu_h} e^{\nu z} \spinHeun{\omega}{z}.
\eeq
In eq.~\eqref{HomTrasfR}, we have re-labelled the coefficients $\mu_i$ ($i=0,1$) from sec.~\ref{sec:HeunCanonical} so that one can explicitly associate them with their respective hypersurfaces, i.e $\mu_c$ relates to the Cauchy horizon at $z=0$, whereas $\mu_h$ to the event horizon at $z=1$. 
Their values, as well as those for $\nu$ follow after substituting eqs.~\eqref{eq:coeff_ADE}-\eqref{eq:coeff_C} into eq.~\eqref{eq:homo_coeff_fix}
\bea
\label{eq:mu_c}
& \mu_c^{+} = - \spin -  \dfrac{\i\,  \omega - \i m \Omega_{\rm c} }{2 \varkappa_{\rm c}}, \quad \mu_c^{-} =   \dfrac{\i\,  \omega - \i m \Omega_{\rm c} }{2 \varkappa_{\rm c}},\\
\label{eq:mu_h}
& \mu_h^{+} =   - \spin -  \dfrac{\i\,  \omega - \i m \Omega_{\rm h} }{2 \varkappa_{\rm h}}, \quad \mu_h^{-} =    \dfrac{\i\,  \omega - \i m \Omega_{\rm h} }{2 \varkappa_{\rm h}}, \\
\label{eq:nu}
& \nu^+ = + \i \,  \rh \omega \,(1- \kappa^2), \quad \nu^- = - \i \,  \rh \omega \,(1- \kappa^2).
\eea
As discussed in sec.~\ref{sec:HeunCanonical}, for the particular choice of coordinate mapping \eqref{eq:z_of_r_BH}, there is a total of 8 combinations for the parameters $\{ \mu_c, \mu_h, \nu\}$, each of which giving rise to a different representation of the radial Teukolsky equation \eqref{eq:RadialTeuk_BL} in the canonical confluent Heun form for the function $\spinHeun{\omega}{z}$. 

Table \ref{tab:HeunCoeff} collects all the 8 possibilities for the homotopic parameters $\{ \mu_c, \mu_h, \nu\}$. It displays only the corresponding values for the Heun coefficients $\gamma, \delta, \epsilon, \alpha$. As mentioned in the introduction, a practical effect of the homotopic transformation \eqref{HomTrasfR} is to modify the asymptotic behaviour for the solutions \eqref{eq:R_in_and_out_BL}. The expression for $q$ follows directly from eq.~\eqref{eq:canonical_par}, but we omit them in the table, as they are slightly lengthier, and they do not affect the characteristic exponents. Indeed, according to eqs.~\eqref{eq:CharExp_rho} and \eqref{eq:CharExp_sigeta}, the resulting characteristic exponents $\tilde \varrho$, $\tilde \eta$ and $\tilde \varsigma$ are determined directly from the canonical Heun coefficients $\gamma, \delta, \epsilon, \alpha$. An alternative practical calculation is to recall that the transformation \eqref{HomTrasfR} effectively modifies the exponents of eq.~\eqref{eq:R_BL_asymp} by
\beq
\label{eq:CharcCoeff_Change}
\tilde \varrho_{{\cal C}_{\rm p/f}} = \varrho_{{\cal C}_{\rm p/f}} - \mu_c, \quad \tilde \varrho_{{\cal H}_{\rm p/f}} = \varrho_{{\cal H}_{\rm p/f}} - \mu_h, \quad \tilde \eta_{\infty^{\pm}} = \eta_{\infty^{\pm}} + \mu_c + \mu_h, \quad \tilde \varsigma_{\infty^\pm} = \varsigma_{\infty^\pm}  - \dfrac{ \nu}{(r_h-r_c)}. 
\eeq
Table \ref{tab:HeunExp} collects the resulting characteristic exponents for the all the 8 configurations.

\smallskip
In the next sections, we provide a spacetime interpretation for all these 8 configurations, and we show that the different  asymptotic behaviour is a consequence of a choice for a different spacetime foliation.

 \begin{table*}[t!]
  \caption{Canonical coefficients of confluent Heun equation for the radial Teukolsky equation.}
\centering 
\begin{tabular}{|c||   c|c|c|c|}
 \hline 
 $(\mu_c, \mu_h, \nu)$ &     $\gamma$ & $\delta$  & $\epsilon $ & $\alpha$ \\
\hline
\hline
  $(-, -, +)$ & $1 + \i \dfrac{\omega - m \Omega_{\rm c}}{\varkappa_c} + \spin$ & $1 + \i\dfrac{\omega - m \Omega_{\rm h}}{\varkappa_h}  + \spin$  & $\dfrac{4 \i  \varkappa_h  a \omega}{\Omega_{\rm h}}$ & $  \dfrac{4 a\omega\, \i\varkappa_h (2\spin +1)}{\Omega_{\rm h}}$ \\
  \hline
   $(+, +, -)$ & $1 - \i \dfrac{\omega - m \Omega_{\rm c}}{\varkappa_c} - \spin$ & $1 - \i\dfrac{\omega - m \Omega_{\rm h}}{\varkappa_h}  - \spin$ & $-\dfrac{4 \i  \varkappa_h  a \omega}{\Omega_{\rm h}}$ & $  \dfrac{4 a\omega\, \i\varkappa_h (2\spin -1)}{\Omega_{\rm h}}$\\
 \hline 
 \hline
 $(-, -, -)$ & $1 + \i \dfrac{\omega - m \Omega_{\rm c}}{\varkappa_c} + \spin$ & $1 + \i\dfrac{\omega - m \Omega_{\rm h}}{\varkappa_h}  + \spin$  & $-\dfrac{4 \i  \varkappa_h  a \omega}{\Omega_{\rm h}}$ & $\dfrac{ 4 a \varkappa_h  \omega}{\rh \Omega_{\rm h}^2}\left( 2 a \omega  -  \i \rh \Omega_{\rm h} \right)$ \\
  \hline
 $(+, +, +)$ & $1 - \i \dfrac{\omega - m \Omega_{\rm c}}{\varkappa_c} - \spin$ & $1 - \i\dfrac{\omega - m \Omega_{\rm h}}{\varkappa_h}  - \spin$ & $\dfrac{4 \i  \varkappa_h  a \omega}{\Omega_{\rm h}}$ & $\dfrac{ 4 a \varkappa_h  \omega}{\rh \Omega_{\rm h}^2}\left( 2 a \omega  +  \i \rh \Omega_{\rm h} \right)$ \\
 \hline
 \hline 
 $(-, +, +)$ & $1 + \i \dfrac{\omega - m \Omega_{\rm c}}{\varkappa_c} + \spin$ & $1 - \i\dfrac{\omega - m \Omega_{\rm h}}{\varkappa_h}  - \spin$ & $\dfrac{4 \i  \varkappa_h  a \omega}{\Omega_{\rm h}}$ & $ - 4 a\omega \left( m - \dfrac{\omega + \i\varkappa_h(\spin +1)}{\Omega_{\rm h}}\right)$\\
 \hline 
  $(+, -, -)$ & $1 - \i \dfrac{\omega - m \Omega_{\rm c}}{\varkappa_c} - \spin$ & $1 + \i\dfrac{\omega - m \Omega_{\rm h}}{\varkappa_h}  + \spin$  & $-\dfrac{4 \i  \varkappa_h  a \omega}{\Omega_{\rm h}}$ & $ - 4 a\omega \left( m - \dfrac{\omega + \i\varkappa_h(\spin -1)}{\Omega_{\rm h}}\right)$ \\
 \hline 
 \hline
 $(-, +, -)$ & $1 + \i \dfrac{\omega - m \Omega_{\rm c}}{\varkappa_c} + \spin$ & $1 - \i\dfrac{\omega - m \Omega_{\rm h}}{\varkappa_h}  - \spin$ & $-\dfrac{4 \i  \varkappa_h  a \omega}{\Omega_{\rm h}}$ & $  4 a\omega \left( m - \dfrac{\omega + \i\varkappa_c(\spin -1)}{\Omega_{\rm c}}\right)$  \\
 \hline
 $(+, -, +)$ & $1 - \i \dfrac{\omega - m \Omega_{\rm c}}{\varkappa_c} - \spin$ & $1 + \i\dfrac{\omega - m \Omega_{\rm h}}{\varkappa_h}  + \spin$  & $\dfrac{4 \i  \varkappa_h  a \omega}{\Omega_{\rm h}}$  & $ 4 a\omega \left( m - \dfrac{\omega + \i\varkappa_c(\spin +1)}{\Omega_{\rm c}}\right)$\\  
  \hline 
\end{tabular}
\label{tab:HeunCoeff}
\end{table*}

\begin{table*}[t!]
  \caption{Characteristic exponents of confluent Heun solution for the radial Teukolsky equation.}
\centering 
\begin{tabular}{|c||   c|c|c|c|c|c|}
 \hline 
 $(\mu_c, \mu_h, \nu)$ &      $\tilde \varrho_{{\cal H}_{\rm p}} $ & $\tilde \varrho_{{\cal H}_{\rm f}}$ & $\tilde \eta_{\infty^{-}}$ & $\tilde \eta_{\infty^{+}}$ & $\tilde \varsigma_{\infty^{-}}$ & $\tilde \varsigma_{\infty^{+}}$ \\
\hline
\hline
  $(-, -, +)$ & $ 0$& $ -\dfrac{\i \omega}{ \varkappa_h} + \dfrac{\i m \Omega_h}{ \varkappa_h} -\spin$& $ 1+ 4 \i M \omega$&$1+2\spin$ & $ - 2 \i \omega $ &  $0$  \\
  \hline
   $(+, +, -)$ & $ \dfrac{\i \omega}{ \varkappa_h} - \dfrac{\i m \Omega_h}{ \varkappa_h} +\spin$& $0$&$1-2 \spin$ &$1- 4 \i M \omega$ & $0$  & $2 \i \omega$    \\
 \hline 
 \hline
 $(-, -, -)$ &$0$ & $ -\dfrac{\i \omega}{ \varkappa_h} + \dfrac{\i m \Omega_h}{ \varkappa_h} -\spin$&$1+4 \i M \omega$  & $1+ 2 \spin$ & $0$  & $2 \i \omega$    \\
  \hline
 $(+, +, +)$ &$ \dfrac{\i \omega}{ \varkappa_h} - \dfrac{\i m \Omega_h}{ \varkappa_h} +\spin$ &$0$ & $1-2 \spin$ &$ 1-4 \i M \omega$ & $ - 2 \i \omega $ &  $0$  \\
 \hline
\hline
 $(-, +, +)$ &$ \dfrac{\i \omega}{ \varkappa_h} - \dfrac{\i m \Omega_h}{ \varkappa_h} +\spin$  & $0$ & $1- \spin + \dfrac{\i \omega - \i m \Omega_c}{ \varkappa_c}$ & $1+ \spin -\dfrac{ \i \omega - \i m \Omega_h}{ \varkappa_h}$ & $ - 2 \i \omega $ &  $0$  \\
 \hline 
  $(+, -, -)$ & $0$ & $ -\dfrac{\i \omega}{ \varkappa_h} + \dfrac{\i m \Omega_h}{ \varkappa_h} -\spin$ & $1- \spin +\dfrac{\i \omega - \i m \Omega_h}{ \varkappa_h}$  &$1+ \spin - \dfrac{ \i \omega - \i m \Omega_c}{ \varkappa_c}$ & $0$  & $2 \i \omega$    \\
 \hline 
 \hline
 $(-, +, -)$ &$ \dfrac{\i \omega}{ \varkappa_h} - \dfrac{\i m \Omega_h}{ \varkappa_h} +\spin$ &$0$ & $1- \spin + \dfrac{\i \omega - \i m \Omega_c}{ \varkappa_c}$ & $1+ \spin -\dfrac{ \i \omega - \i m \Omega_h}{ \varkappa_h}$ & $0$  & $2 \i \omega$    \\
 \hline
 $(+, -, +)$ & $0$ & $ -\dfrac{\i \omega}{ \varkappa_h} + \dfrac{\i m \Omega_h}{ \varkappa_h} -\spin$ & $1- \spin +\dfrac{\i \omega - \i m \Omega_h}{ \varkappa_h}$  &$1+ \spin - \dfrac{ \i \omega - \i m \Omega_c}{ \varkappa_c}$ & $ - 2 \i \omega $ &  $0$  \\
  \hline 
\end{tabular}
\label{tab:HeunExp}
\end{table*}

\section{A spacetime interpretation to the Confluent Heun Equation solutions}\label{sec:CHE_Spacetime}
We now present an alternative strategy to obtain the radial Teukolsky equation in the canonical form for the CHE. Instead of employing s-homotopic transformations to the radial ODE, we focus on the spacetime geometry, with particular attention to the role played by the choice of time coordinate. To exemplify the strategy, we begin discussing the well-know representation of the Teukolsky equation in Kerr coordinates \cite{Teukolsky:1974yv}, and then we consider generic time slices.

\subsection{Outgoing and Ingoing Kerr coordinates}\label{sec:OutIn}
Let us first introduce the outgoing Kerr coordinates $\hat x^\mu = \OutNullCoord$ via 
\beq
\label{eq:OutKerr}
t = u +  r^*(r), \qquad  \phi = \hat \phi + \chi(r),
\eeq
with the tortoise coordinate $r^*(r)$ and phase $\chi(r)$ defined in eqs.~\eqref{eq:tort_r} and \eqref{eq:tort_chi}. The resulting line element is not directly relevant to the discussion, but the expressions for the Kinnersley null tetrad \eqref{eq:Kin_null_tetrad_lk} plays an important role as they define the Weyl curvature components $\Psi_0$ and $\Psi_4$. In the coordinate system $\hat x^\mu$, the outgoing and ingoing Kinnersley null vectors read
\beq
\ell^{\hat \mu}_+ = \delta^{\hat \mu}_r, \qquad \ell^{\hat \mu}_- = \dfrac{1}{\Sigma(r,\theta)} \left( (r^2 + a^2) \delta^{\hat \mu}_u   -\dfrac{\Delta(r)}{2}\delta^{\hat \mu}_r + a \delta^{\hat \mu}_{\hat\phi}   \right).
\eeq
As expected, the Kinnersley null tetrad is adapted to the outgoing Kerr coordinates $\hat x^a = \OutNullCoord$, and the vectors' components are well-defined at the horizons, where $\Delta(r) = 0$. Moreover, when $\Delta(r) = 0$, the ingoing null vector $\ell^a_-$ is the generator of the horizon surfaces, indicating that $\rh$ corresponds to the white hole horizon.

Because the null tetrad is adapted to the coordinate system, one can identify the Teukolsky master function in outgoing Kerr coordinates $\spinOutNullPsi{\spin}(\hat x^a)$ directly with the master function in Boyer-Lindquist coordinates via
\beq
\label{eq:TeukFunc_BL_OutKerr}
\spinOutNullPsi{\spin}(\hat x^a) = \spinPsi{\spin}(x^a),
\eeq
where the transformation $x^a(\hat x^a)$ between Boyer-Lindquist and outgoing Kerr coordinates is understood.

\smallskip
An equivalent procedure applied to eq.~\eqref{eq:Teukolsky_BL} leads to the following form of the Teukolsky equation 
\bea
\label{eq:WaveTeuk_OutKerr}
0 = &a^2 \sin^2\theta\, \spinOutNullPsi{\spin}_{,uu} (\hat x^a)   + 2 a\, \left( \spinOutNullPsi{\spin}_{,u \hat \phi} (\hat x^a) - \spinOutNullPsi{\spin}_{,r \hat \phi}(\hat x^a)  \right) - 2(r^2 + a^2)\, \spinOutNullPsi{\spin}_{,ur} (\hat x^a)+ \nn \\
& - 2 \bigg( \i a \spin \cos\theta + r (1+2\spin)\bigg)\spinOutNullPsi{\spin}_{,u}(\hat x^a) + \Delta^{-\spin}(r) \, \partial_r \bigg( \Delta^{\spin+1} (r)\, \spinOutNullPsi{\spin}_{,r}(\hat x^a)  \bigg) + \\
&+ \dfrac{1}{\sin\theta}\partial_\theta \bigg( \sin\theta \,  \spinPsi{\spin}_{,\theta}(\hat x^a) \bigg) + \dfrac{1}{\sin^2\theta} \, \spinOutNullPsi{\spin}_{,\hat \phi \hat \phi} (\hat x^a) +2\i \spin \dfrac{\cos\theta}{\sin^2\theta} \, \spinOutNullPsi{\spin}_{,\hat \phi}(\hat x^a)  - \spin(\spin\cot^2\theta -1) \,\spinOutNullPsi{\spin}(\hat x^a). \nn
\eea
Proceeding now to the decomposition into the frequency domain, we introduce
\beq
\label{eq:Fourier_OutKerr}
\spinOutNullPsi{\spin}\OutNullCoord = \dfrac{1}{2 \pi} \int_{-\infty }^{\infty } \sum_{\ell = |\spin|}^{\infty}\sum_{m=-\ell}^{\ell} e^{-\i \omega u} e^{\i m \hat \phi} \spinOutNullR{\omega} {r}\spinS{\omega} {\rm d} \omega,
\eeq
which yields the radial equation
\bea
\label{eq:RadialEq_OutKerr}
& \Delta^{-\spin}(r) \, \dfrac{d}{dr} \bigg( \Delta^{\spin+1}(r) \,  \dfrac{d}{dr} \bigg) \spinOutNullR{\omega} {r} + 2 \i \bigg(  \omega (r^2 + a^2)\dfrac{d}{dr} -  am   \bigg)  \dfrac{d}{dr}  \spinOutNullR{\omega} {r} \\ 
&- \bigg( A_{\ell m } + a^2\omega^2 - 2 a \omega m - 2 \i r\omega (1+2\spin) \bigg) \spinOutNullR{\omega} {r} = 0. \nn
\eea
Upon division of eq.~\eqref{eq:RadialEq_OutKerr} by $\Delta(r)$, a direct comparison against the generic CHE \eqref{eq:GenConfHeun} shows that eq.~\eqref{eq:RadialEq_OutKerr} is already in the canonical form, i.e., with the corresponding coefficients $\hat{B}_i = \hat{D} =0$. The remaining coefficients read
\bea
\label{eq:CHE_Coeff_OutKerr_AE}
&\hat A_0 = 1 + \i \dfrac{\omega - m \Omega_{\rm c}}{\varkappa_c} + \spin, \quad \hat  A_1 = 1 + \i\dfrac{\omega - m \Omega_{\rm h}}{\varkappa_h}  + \spin, \quad \hat E =  2 \i \omega, \\
&\hat C_0 = \dfrac{\Omega_{\rm c}}{2 a \varkappa_{\rm c} }\bigg( -A_{\ell m} + 2m a \omega - a^2 \omega^2 + 2\i \,a \omega \kappa (1+ 2\spin) \bigg), \\ 
\label{eq:CHE_Coeff_OutKerr_C1}
&\hat C_1 = \dfrac{\Omega_{\rm h}}{2 a \varkappa_{\rm h} }\bigg( -A_{\ell m} + 2m a \omega - a^2 \omega^2 + 2\i \, \rh \omega (1+ 2\spin) \bigg).  
\eea
The canonical coefficients $(\gamma, \delta, \epsilon, \alpha, q)$ then follow simply from the coordinate transformation \eqref{eq:z_of_r_BH}, with the resulting canonical CHE expressed in terms of the function
\beq
\label{eq:HeunR_+-+}
\spinHeunType{\omega}{z}{\scriptscriptstyle{(--+)}} = \spinOutNullR{\omega} {r(z)}.
\eeq 
Eq.~\eqref{eq:HeunR_+-+} emphasises two features. Firstly as explained, no s-homotopic transformation is needed when the radial equation results from the original Teukolsky equation expressed in outgoing Kerr coordinates. Secondly, the notation $\spinHeunType{\omega}{z}{\scriptscriptstyle{(--+)}}$ indicates that the Heun function relates to the combination $(\mu_c^-, \mu_h^-, \nu^+)$ in Table \ref{tab:HeunCoeff}, where one can read the corresponding canonical Heun coefficients $(\gamma, \delta, \epsilon, \alpha, q)$.

To verify that the s-homotopic transformation \eqref{HomTrasfR}, with the particular combination $(\mu_c^-, \mu_h^-, \nu^+)$, relates to functions defined along the hypersurface $u=$constant, we observe that the Fourier transforms \eqref{eq:Fourier_BL} and \eqref{eq:Fourier_OutKerr}, applied to eq.~\eqref{eq:TeukFunc_BL_OutKerr} yields
\bea
& \underbrace{e^{-\i \omega t}}_{e^{-\i \omega(u + r^*) }} \, \underbrace{e^{\i m \phi}}_{ e^{\i m (\hat \phi +\chi)} }\, \spinR{\omega} {r}  =e^{-\i \omega u} e^{\i m \hat \phi} \spinOutNullR{\omega} {r} \nn \\
&  \spinR{\omega} {r} = \hat {\cal Z}(r) \spinOutNullR{\omega} {r}, \quad \hat {\cal Z}(r) = e^{\i \omega  r^*(r)} \, e^{- \i m \chi(r)}.
\eea
The function $\hat {\cal Z}(r)$ arises as a natural consequence of the change of time and angular coordinates when going from BL coordinates $x^a$ to outgoing Kerr coordinates $\hat x^a$. When considering the radial re-scaling \eqref{eq:z_of_r_BH}, it reads 
\beq
\label{eq:Z_OutKerr}
\hat {\cal Z}(r(z)) \propto e^{\i \omega (\rh - \rc) z} \, (z-1)^{i (\omega - m \Omega_{\rm h})/(2\varkappa_{\rm h})}\, z^{i (\omega - m \Omega_{\rm c})/(2\varkappa_{\rm c})}.
\eeq 
Thus, eq.~\eqref{eq:Z_OutKerr} predicts exactly the values $(\mu_c^-, \mu_h^-, \nu^+)$ to input in eq.~\eqref{HomTrasfR} from a pure spacetime argument via the change of time and angular coordinates.

\smallskip
A very similar argument follows if we consider the ingoing Kerr coordinates $\check x^a = \InNullCoord$, defined by
\beq
\label{eq:InKerr}
t = v -  r^*(r), \quad  \phi = \check \phi - \chi(r).
\eeq
In this case, however, the Kinnersly tetrad is not best suited to define the scalars $\Psi_0$ and $\Psi_4$. In coordinates $\check x^a$, the outgoing and ingoing Kinnersley null vectors read
\beq
\ell^{\check \mu}_+ =  \dfrac{2 (r^2 + a^2)}{\Delta(r)} \delta^{\check \mu}_v   + \delta^{\check \mu}_r + \dfrac{2a}{\Delta(r)} \delta^{\check \mu}_{\check\phi}   ,  \quad \ell^{\check \mu}_- = -\dfrac{\Delta(r)}{2\Sigma(r, \theta)} \delta^{\check \mu}_r, 
\eeq
and hence, ill-defined at the horizons, when $\Delta(r) = 0$. Therefore, a new set of null tetrad is needed, boosted as
\beq
\label{eq:boost_null_tetrad}
\check \ell^a_+ = \Delta(r)\,  \ell^{\check a}_+, \quad \check \ell^a_- = \Delta^{-1}(r) \ell^{\check a}_-.
\eeq
Apart from being regular at the horizons, one can infer that the surface $\rh$ now corresponds to the black hole horizon because the outgoing null vector $\check \ell^a_+$ is the generator of the horizon.

The boost transformation \eqref{eq:boost_null_tetrad} implies a change in the curvature scalars $\check \Psi_0(\check x^\mu) = \Delta^2 (r)\Psi_0(x^\mu)$ and $\check \Psi_4(\check x^\mu) = \Delta^{-2} (r)\Psi_4(x^\mu)$, suggesting a suitable re-scaling of the Teukolsky master function via
\beq
\label{eq:InKerr_MasterFunc}
\spinInNullPsi{\spin}(\check x^a) = \Delta^{\spin}(r)\, \spinPsi{\spin}(x^a).
\eeq
Now, the same arguments follow as in the case of outgoing Kerr coordinates. The function $\spinInNullPsi{\spin}(\check x^a)$ satisfies a wave equation similar to \eqref{eq:WaveTeuk_OutKerr}, with $\partial_{u} \rightarrow -\partial_{v}$, $\partial_{\hat \phi} \rightarrow -\partial_{\check \phi}$ and $\spin \rightarrow - \spin$. Similarly, a frequency domain decomposition with respect to $v$ and $\check \phi$, yields a radial equation for a variable $\spinInNullR{\omega}{r}$ related to eq.~\eqref{eq:RadialEq_OutKerr} via $\omega\rightarrow -\omega$, $m\rightarrow -m$ and $\spin \rightarrow - \spin$. Besides, the same symmetries apply to the CHE coefficients \eqref{eq:CHE_Coeff_OutKerr_AE}-\eqref{eq:CHE_Coeff_OutKerr_C1}, and the canonical Heun coefficients resulting from simple coordinate re-scaling
\beq
\spinHeunType{\omega}{z}{\scriptscriptstyle{(++-)}} = \spinInNullR{\omega} {r(z)}.
\eeq
From the symmetry between outgoing and ingoing Kerr coordinates, a direct relation between the function $\spinInNullR{\omega} {r}$ with the configuration $(\mu_c^+, \mu_h^+, \nu^-)$ in Table \ref{tab:HeunCoeff} might have been expected. However, eqs.~\eqref{eq:mu_c} and \eqref{eq:mu_h} are not exactly symmetric under a sign change, as $\mu^+_c$ and $\mu^+_h$ pick a factor of $\spin$, absent in their counterparts $\mu^-_c$ and $\mu^-_h$. The spacetime approach provides an understanding of the terms with $\spin$ via the boost re-scaling \eqref{eq:boost_null_tetrad}. Indeed, upon a frequency domain decomposition, the original eq.~\eqref{eq:InKerr_MasterFunc} implies
\beq
\spinR{\omega} {r} = \check {\cal Z}(r) \spinInNullR{\omega} {r}, \quad \check {\cal Z}(r) = \Delta^{-\spin}(r) e^{-\i \omega  r^*(r)} \, e^{+ \i m \chi(r)},
\eeq
with
\beq
\label{eq:Z_InKerr}
\check {\cal Z}(r(z)) \propto e^{-\i \omega (\rh - \rc) z} \, (z-1)^{-\spin-i (\omega - m \Omega_{\rm h})/(2\varkappa_{\rm h})}\, z^{-\spin-i (\omega - m \Omega_{\rm c})/(2\varkappa_{\rm c})}.
\eeq
Thus, not only does eq.~\eqref{eq:Z_InKerr} predict the values $(\mu_c^+, \mu_h^+, \nu^-)$ from the change of time and angular coordinates, but it also incorporates the boost re-scaling necessary to make the function regular at the black hole horizon. Most importantly, an analysis simply based on the s-homotopic transformation \eqref{HomTrasfR} does not distinguish the spacetime meaning of the surface $\rh$ as corresponding either to the white hole horizon in configuration $(--+)$ or the black hole horizon in configuration $(++-)$.

\subsection{Heun slices}\label{sec:Heun Slices}
The previous section has shown how the configurations $(\mu_c^-, \mu^-, \nu^+)$ and $(\mu_c^+, \mu^+, \nu^-)$ --- originally derived from the s-homotopic transformation of radial Teukolsky function \eqref{HomTrasfR} --- are actually a direct consequence of changing the spacetime time coordinate from the Boyer-Lindquist $t$ into outgoing and ingoing Kerr coordinates $u$ and $v$, respectively. 

While these two time coordinates are well-known in the literature, the existence of 8 possible configurations $(\mu_c, \mu, \nu)$ for homotopic transforming the radial Teukolsky equation into the CHE canonical form suggests that further spacetime transformations might be of relevance. In this section, we study the geometrical structure of these 8 coordinate transformations and we shall refer to the resulting spacetime hypersurfaces as Heun slices.

\smallskip
We begin by introducing a generic coordinate system $\tilde x^\mu = \GenCoord$ related to the BL coordinates via
\beq
\label{eq:GenCoordSys}
t = \t - h(r), \quad \phi = \tilde \phi - \xi(r). \quad  
\eeq
As discussed in the example for the ingoing Kerr coordinate $\check x^\mu$, the resulting components of the outgoing and ingoing Kinnersley tetrads $\ell^a_+$ and $\ell^a_-$ might not be well-defined at the horizons $r=\rh$ and/or $r=\rc$. Therefore, associated with the coordinate change \eqref{eq:GenCoordSys}, we also allow for a possible boost to the Kinnersley tetrads via
\beq
\label{eq:GenTetrad}
\tilde \ell^a_+ = \zeta(r) \ell^a_+, \quad \tilde \ell^a_- = \zeta(r)^{-1} \ell^a_-.
\eeq
Eqs.~\eqref{eq:GenCoordSys} and \eqref{eq:GenTetrad} imply a transformation for the Teukolsky master function in the form  
\beq
\label{eq:InKerr_MasterFunc}
\spinGenPsi{\spin}(\tilde x^\mu) = \zeta^{\spin}(r) \spinPsi{\spin}(x^\mu).
\eeq
Upon a decomposition into the frequency domain via
\beq
\label{eq:Fourier_Gen}
\spinGenPsi{\spin}\GenCoord = \dfrac{1}{2 \pi} \int_{-\infty }^{\infty } \sum_{\ell = |\spin|}^{\infty}\sum_{m=-\ell}^{\ell} e^{-\i \omega \t} e^{\i m \tilde \phi} \spinGenR{\omega} {r}\spinS{\omega} {\rm d} \omega,
\eeq
the new radial function $\spinGenR{\omega} {r}$ relates to the original $\spinR{\omega} {r}$ by
\beq
\label{eq:GenRadialTrasfo}
\spinR{\omega} {r} = \tilde {\cal Z}(r)\, \spinGenR{\omega} {r}, \quad \tilde {\cal Z}(r) = \zeta(r)^{-\spin}\, e^{-\i \omega h(r)} e^{\i m \zeta(r)}
\eeq
A direct comparison of eq.~\eqref{eq:GenRadialTrasfo} against the homotopic transformation \eqref{HomTrasfR} allows us to identify the functions $h(r)$ and $\xi(r)$ for each one of the 8 possible configurations in Table \ref{tab:HeunCoeff}. Such a comparison also provides the boost function $\zeta(r)$, with the regularity of the null tetrads \eqref{eq:GenTetrad} at the horizons being a consistency check of the result.

Motivated by our previous knowledge that $h(r) = \pm r^*(r)$, $\xi(r) = \pm \chi(r)$ for the outgoing and ingoing Kerr coordinates, with also $\zeta(r) = \Delta(r)$ in the latter, we consider an Ansatz for the functions $h(r)$, $\xi(r)$ and $\zeta(r)$ as modifications of $r^*(r)$,  $\chi(r)$ and $\Delta(r)$ in the form
\bea
\label{eq:Height_HeunSlices}
&&h(r) =a_\infty \, r + 2 M\, b_\infty  \ln\left( \dfrac{r}{\rh}\right) + \dfrac{b_h}{2 \varkappa_{\rm h}}  \ln\left( 1- \dfrac{\rh}{r}\right) +  \dfrac{b_c}{2 \varkappa_{\rm c}}  \ln\left( 1- \dfrac{\rc}{r}\right), \\
&&\xi(r) = c_\infty \ln \left( \dfrac{r}{\rh} \right) + c_h \dfrac{ \Omega_{\rm h}}{2 \varkappa_{\rm h}}  \ln\left( 1- \dfrac{\rh}{r}\right) + c_c \dfrac{ \Omega_{\rm c}}{2 \varkappa_{\rm c}}  \ln\left( 1- \dfrac{\rc}{r}\right), \\
\label{eq:boost_zeta}
&&\zeta(r)  = (r-\rh)^{d_h} (r-\rc)^{d_c},
\eea
Table \ref{tab:HeunCoeffSpacetime} collects all the coefficients $a$'s, $b$'s, $c$'s and $d$'s from the above Ansatz. Note that the coefficients satisfy the constraint
\beq
\label{eq:constraint_coeff}
2 M b_\infty = \dfrac{b_h}{2\varkappa_h} + \dfrac{b_c}{2\varkappa_c}, \quad c_\infty = \dfrac{c_h \Omega_h}{2\varkappa_h} + \dfrac{c_c  \Omega_c}{2\varkappa_c},
\eeq
which follows from the identity of the s-homotopic coefficients $$(r-\rc)^{\mu_c} (r-\rh)^{\mu_h} = r^{\mu_c + \mu_h} (1-\rc/r)^{\mu_c} (1-\rh/r)^{\mu_h}.$$

\begin{table*}[h!]
  \caption{Coefficients for functions $h(r), \xi(r)$ and $\zeta(r)$ associated with the spacetime interpretation for the Confluent Heun function}
\centering 
\begin{tabular}{|c||   c|c|c|c|| c|c|c|| c|c|}
 \hline 
 $(\mu_c, \mu_h, \nu)$ &     $a_\infty$ & $b_\infty$  & $b_h $ & $b_c$ & $c_\infty$ &$c_h$ & $c_c$ &$d_h$ & $d_c$ \\
\hline
\hline
  $(-, -, +)$ & $-1$ & $-1$ & $-1$ & $-1$ & 0 & $-1$ & $-1$ & $0 $ & $0$ \\
  \hline
$(+, +, -)$ & $+1$ & $+1$ & $+1$ & $+1$ & 0 & $+1$ & $+1$ & $1$ & $1$ \\
\hline 
 \hline
 $(-, -, -)$ & $+1$ & $-1$ & $-1$ & $-1$ & 0 & $-1$ & $-1$ & $0 $ & $0$ \\
  \hline
 $(+, +, +)$ & $-1$ & $+1$ & $+1$ & $+1$ & 0 & $+1$ & $+1$ & $1$ & $1$ \\
  \hline
 \hline 
 $(-, +, +)$ & $-1$ & $  +\left( 2 \rh \varkappa_{\rh} \right)^{-1} $ & $+1$ & $-1$ & $+\Omega_h/\varkappa_h$ & $+1$ & $-1$ & $1$  & $0$    \\
 \hline 
  $(+, -, -)$ & $+1$ & $  -\left( 2 \rh \varkappa_{\rh} \right)^{-1} $ & $-1$ & $+1$ & $-\Omega_h/\varkappa_h$ & $-1$ & $+1$  & $0 $ & $1$ \\
 \hline 
 \hline
 $(-, +, -)$ &  $+1$ & $  +\left( 2 \rh \varkappa_{\rh} \right)^{-1} $ & $+1$ & $-1$ & $+\Omega_h/\varkappa_h$ & $+1$ & $-1$ & $1$  & $0$    \\ 
 \hline
 $(+, -, +)$ &  $-1$ & $  -\left( 2 \rh \varkappa_{\rh} \right)^{-1} $ & $-1$ & $+1$ & $-\Omega_h/\varkappa_h$ & $-1$ & $+1$ & $0 $ & $1$ \\
  \hline 
\end{tabular}
\label{tab:HeunCoeffSpacetime}
\end{table*}

We will refer to each surface resulting from the 8 configurations as \emph{Heun slice}. To study the Heun slices, we employ the following geometrical measures to infer the behaviour of each $\t =$constant surface. 

\subsubsection*{Heun slices in the Carter-Penrose diagram of Kerr spacetimes}
We depict the surfaces $\t =$ constant in the Carter-Penrose diagram, restricted to one block element between the past and future singularity surfaces at $r=0$. To the past, the white hole region is delimited by the diamond with boundaries ${\cal C}_{\rm p}$ and ${\cal H}_{\rm p}$, representing the coordinate surfaces $\rc$ and $\rh$, respectively. To the future, the same coordinates values will delimit the black hole region within the surfaces ${\cal C}_{\rm f}$ and ${\cal H}_{\rm f}$. 

The Heun slices are calculated and drawn within the diagram according to the following transformation from the coordinates $\GenCoord$ \eqref{eq:GenRadialTrasfo} at $\theta=0$ into Penrose coordinates $(T,R)$ via
\beq
T(U,V)= - \arctan(U(u))+ \arctan(V(v)), \qquad R(U,V)=\arctan(U(u))+ \arctan(V(v)),
\eeq
with $(U,V)$ Kruskal-Szekeres coordinates defined in terms of the outgoing and ingoing Kerr coordinates via
\beq
U(u) = \left\{
\begin{array}{cc}
+\exp\bigg{[}-\varkappa_{\rm h }\, u(\tilde t, r)\bigg{]} & {\rm Exterior} \; {\rm region},\\
-\exp\bigg{[}-\varkappa_{\rm h }\, u(\tilde t, r)\bigg{]}&  {\rm Black} \; {\rm Hole} \; {\rm region},\\
+\exp\bigg{[}-\varkappa_{\rm h }\, u(\tilde t, r)\bigg{]} &  {\rm White} \; {\rm Hole} \; {\rm region},
\end{array}
\right.
\eeq
and
\beq
V(v) =  \left\{
\begin{array}{cc}
+\exp\bigg{[}\varkappa_{\rm h }\,v(\tilde t, r)\bigg{]} & {\rm Exterior} \; {\rm region},\\
+\exp\bigg{[}\varkappa_{\rm h }\,v(\tilde t, r)\bigg{]}& {\rm Black} \; {\rm Hole} \; {\rm region},\\
-\exp\bigg{[}\varkappa_{\rm h }\,v(\tilde t, r)\bigg{]} & {\rm White} \; {\rm Hole} \; {\rm region}.\\
\end{array}
\right.
\eeq
Finally, $(u,v)$ relate directly to $(\tilde t, r)$ via
\bea
u(\tilde t, r) &=& t(\tilde t, r) - r_*(r)  = \tilde t - h(r) - r_*(r); \\
 v(\tilde t, r) &=& t(\tilde t, r) + r_*(r) = \tilde t - h(r) + r_*(r),
\eea
with the second expression following from  eq.~\eqref{eq:GenCoordSys}. In particular, one see directly that for the $(--+)$ case, $h(r) = -r_*(r)$, and therefore $u = \tilde t$, or alternatively for the $(++-)$ case $h(r) = r_*(r)$, which yields $v = \tilde t$.

\smallskip

\begin{figure*}[t]
\includegraphics[width=0.5\textwidth]{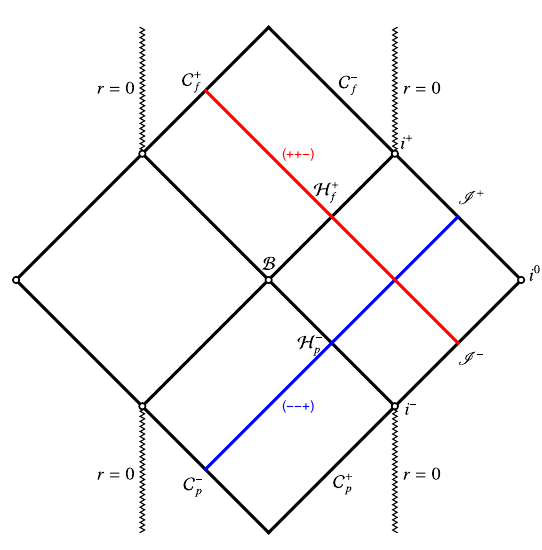}
 \includegraphics[width=0.5\textwidth]{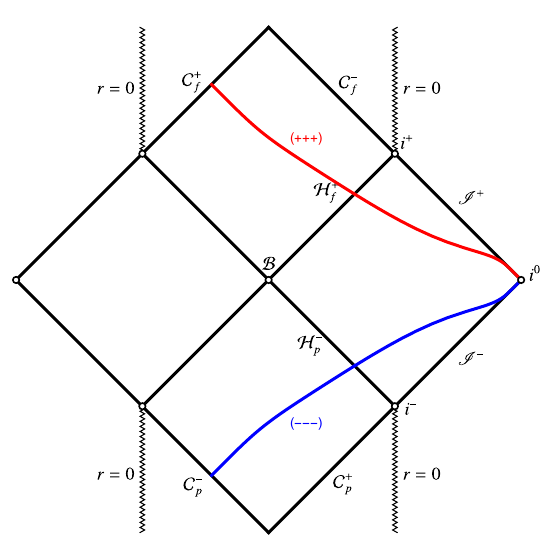}
 \includegraphics[width=0.5\textwidth]{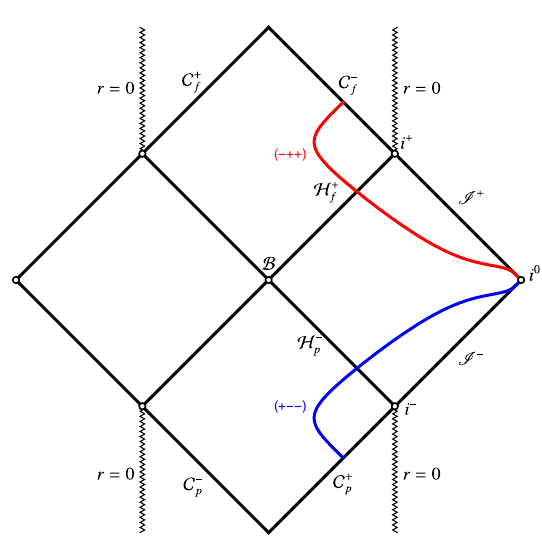}
 \includegraphics[width=0.5\textwidth]{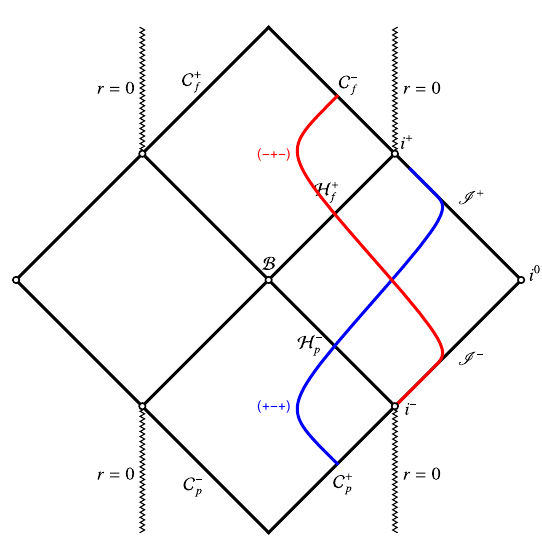}
  \caption{Carter-Penrose diagrams representing the $\t = 0$ hypersurfaces corresponding to the $8$ possible combinations of $(\mu_c^\pm, \mu_h^\pm, \nu^\pm)$ yielding a Heun slice. All configurations are horizon penetrating, with the particular sign in $\mu_c^\pm$ and $\mu_h^\pm$ indicating if the horizon is generated by the outgoing $\tilde \ell^a_+$ or ingoing  $\tilde \ell^a_-$ null vector. The geometrical meaning of $r\rightarrow \infty$ along $\t = \text{constant}$ differs according to each configuration: it may represent future or past null infinity (top left), spatial infinity (top right and bottom left), or even timelike infinity (bottom right).}
    \label{fig:penrose_diagrams}
\end{figure*}

Figure~\ref{fig:penrose_diagrams} consists of 4 panels depicting an example of $\t=$ constant hypersurface for configuration pairs related to each other by a time inversion within one asymptotic flat region. A sign $+/-$ associated to the choice $\mu_c^\pm$ and $\mu_h^\pm$ is also assigned to each of the horizons's past and future surfaces ${\cal C}_{\rm p/f}^{\pm}$ and ${\cal H}_{\rm p/f}^{\pm}$. 

We observe a rich structure, with the limit $r\rightarrow \infty$ along the $\tilde t = \text{constant}$ hypersurfaces possibly representing null, spatial or even timelike infinity ---see Figure \ref{fig:penrose_diagrams}. All these surfaces are horizon penetrating both at the event and Cauchy horizons, but as $r\rightarrow \rc$, the slices $\tilde t = \text{constant}$ may cross a Cauchy horizon generated by either $\tilde \ell_+^a$ or $\tilde \ell_-^a$. 

To study these properties from a more robust geometrical perspective, we also consider the behaviour of the null tetrad at the horizons, and the asymptotic behaviour of the constant mean curvature associated with $\t = \text{constant}$, as discussed below.

\subsubsection*{Outgoing and Ingoing null vectors as horizon generators}

The boost transformation \eqref{eq:GenTetrad}, with $\zeta(r)$ as in eq.~\eqref{eq:boost_zeta} and values according to Table \ref{tab:HeunCoeffSpacetime}, ensures that the outgoing $\tilde \ell^a_+$ and ingoing null tetrad $\tilde \ell^a_-$ are regular at $r=\rc$ and $r=\rh$. When evaluated at a given horizon surface $\rc$ or $\rh$, either $\tilde \ell^a_+$ or $\tilde \ell^a_-$ will be generating the horizon hypersurface. In practical terms, a null vector, say $\tilde \ell^a_-$, is a horizon generator if $\tilde \ell^{\t}_-\neq 0$, $\tilde \ell^{r}_-= 0$ and $\tilde \ell^{\t}_+= 0$, $\tilde \ell^{r}_+ \neq 0$ at the surface. Equivalently, $\tilde \ell^{r}_-\neq 0$, $\tilde \ell^{t}_-= 0$ and $\tilde \ell^{\t}_+ \neq 0$, $\tilde \ell^{r}_+ = 0$ if $\tilde \ell^a_+$ is the horizon generator.

For $r=\rh$, the event horizon surfaces to the past (white hole ${\cal H}_{\rm p}$) and to the future (black hole ${\cal H}_{\rm f}$) are generated, respectively, by the ingoing $\tilde \ell^a_-$ or outgoing null vector $\tilde \ell^a_+$. Since we are restricting ourselves to one asymptotic flat region in the conformal diagram, all configurations for the Heun slices uniquely identify the white hole horizon with $\mu_h^{-}$ and the black hole horizon with $\mu_h^{+}$, i.e., ${\cal H}_{\rm p}^-$ and ${\cal H}_{\rm f}^+$, respectively. Therefore, the sign $\pm$ in the parameter $\mu_h^{\pm}$ identifies the particular null vector $\tilde \ell^a_{\pm}$ generating the event horizon, and as a consequence, the past/future character of the surfaces ${\cal H}$. The same reasoning applies to the Cauchy horizons ${\cal C}$ at $r=\rc$, where the sign $\pm$ in the parameter $\mu_c^{\pm}$ also identifies the particular null vector $\tilde \ell^a_{\pm}$ generating them. However, they do not determine the past and future character of the surfaces ${\cal C}$. 

We can infer a key property of the Heun functions from the fact that all Heun slices are horizon-penetrating. The characteristic exponents associated with the local solutions around the regular singular points always include one value of $\varrho_i = 0$ --- cf. \eqref{eq:CharExp_rho} for the theoretical prediction and Table \ref{tab:HeunExp} for the Teukolsky equation. From a spacetime perspective, the finiteness of the Heun functions at the singular points directly results from the underlying time-constant hypersurface being horizon-penetrating. The behaviour arising from the other characteristic exponent, $\varrho_i \neq 0$, links the second linearly independent local solution to geometrical information associated with the respective time-symmetric horizon surface, either in the future or the past.


\subsubsection*{Hypersurface signature and asymptotic behaviour}
The vector normal to the hypersurface $\t=$ constant has the norm $|| \nabla_a \t ||^2 = g^{\t \t}$. Its sign indicates whether the surface is null $(g^{\t \t}=0)$, timelike $(g^{\t \t}>0)$ or spacelike $(g^{\t \t}<0)$. We find that the Heun timelike surfaces may change signature in the domain $r\in[\rc, \infty)$ and $\theta\in[0, \pi]$. Analysing these features in detail goes beyond the scope of this work, as we are particularly interested in the asymptotic behaviour $r\rightarrow \infty$ along $\t =$ constant.

To study the asymptotic structure of the slice $\t =$ constant, we first consider the signature of the slice in the asymptotic region via a conformal representation of the Kerr spacetime as $\tilde g_{ab} = \Omega^2 g_{ab}$, with $\Omega = 1/r$. Then the {\em conformal} normal vector has the norm
\beq
\tilde N = r^2 g^{\t \t}, 
\eeq
and the limit $r \rightarrow \infty$ will indicate if the surface is asymptotically null $(\tilde N = 0)$, time-like $(\tilde N > 0)$ or spacelike $(\tilde N < 0)$. Besides, we also monitor the limit $r\rightarrow \infty$ of the extrinsic mean curvature
\beq
K(r,\theta) = -\nabla_a n^a, \quad n_a = \dfrac{\nabla_a \t}{\sqrt{|g^{\t \t}|}}.
\eeq
Surfaces approaching $\scri^\pm$ behave as $K(r,\theta) \sim \mp K_0(\theta) + {\cal O}(1/r)$, with $K_0(\theta)>0$. On the other hand, $K \rightarrow 0$ shows that $\t =$ constant is approaching either spatial infinity $i^0$ (spacelike hypersurface), or past and future timelike infinity $i^\pm$ (timelike hypersurface). Furthermore, past or future timelike infinity are distinguishable by the sign $K(r,\theta) \sim \mp 1/\sqrt{r}$ as $r\rightarrow \infty$. Secs.~\ref{sec:--+ / ++- NewNot}-\ref{sec:-+- / +-+ NewNot} discuss these properties for each Heun slice.

\subsubsection{Configuration $(--+)$ and $(++-)$}\label{sec:--+ / ++- NewNot}
These configurations have already been introduced in sec.~\eqref{sec:OutIn} and they correspond to the well-known outgoing and ingoing Kerr coordinates, respectively. The coefficients in Table \eqref{tab:HeunCoeffSpacetime} express the trivial: the values $a_\infty = b_\infty = b_h = b_c = c_h = c_h = \mp 1$ and $c_\infty=0$ just confirm that the coordinate transformation \eqref{eq:GenCoordSys} read $h(r) = \mp r^*(r)$ and $\xi(r) = \mp \chi(r)$ as in eqs.~\eqref{eq:OutKerr} and \eqref{eq:InKerr}.

Interestingly, even though these coordinates align with outgoing or ingoing null directions, the surfaces $u=$ constant or $v =$constant {\em are not} null (except along $\sin \theta = 0$, or in Schwarzschild $a=0$). Though usually understated in the literature, this property is well-known with recent works still discussing the definition of proper null coordinates that are naturally adapted to the horizons and future null infinity of Kerr spacetime~\cite{Arganaraz:2021fpm}.

Indeed, their respectively conformal norm reads $\tilde N = r^2 a^2 \sin^2\theta/(\rh^2 \Sigma(r; \theta)) > 0$. Hence, unless $\sin \theta = 0$, or $a=0$, these surfaces are actually {\em timelike}, all the way until $r\rightarrow \infty$, where $\left.\tilde N\right|_{\infty} = a^2 \sin^2\theta/\rh^2$. The asymptotic behaviour of the extrinsic mean curvature also confirms the expected results: $\left. K \right|_{\infty} <0 $ for $u=$ constant, and $\left. K \right|_{\infty} > 0 $ for $v=$ constant, since these surfaces approach $\scri^+$ and $\scri^-$, respectively.

The left top panel of Figure~\ref{fig:penrose_diagrams} shows these slices, with the well-known structure for the outgoing and ingoing Kerr coordinates: outgoing Kerr coordinates cross the past event horizon ${\cal H}^-_{\rm p}$ and extend towards future null infinity $\scri^+$, whereas ingoing Kerr coordinates extend from past null infinity $\scri^-$ towards the the future event horizon ${\cal H}^+_{\rm p}$.

A similar interpretation between the characteristic exponents \eqref{eq:CharExp_rho} and the geometrical properties of the $\t = \text{constant}$ hypersurfaces, previously discussed for the regular singular points representing the horizons, can also be drawn for the irregular singular point $r \rightarrow \infty$. For the configuration $(-++)$, the peeling properties\footnote{Recall that the characteristic exponents relate to the Teukolsky master function $\spinGenPsi{-2}$, which includes an additional decay factor $\sim r^{4}$ compared to $\Psi_4$, cf.~\eqref{eq:TeukMaster_NewmanPenrose}.} of $\Psi_0 \sim r^{-5}$ and $\Psi_4 \sim r^{-1}$ along $\scri^+$ can be derived from the coefficients $( \tilde \eta_{\infty^+} , \tilde \varsigma_{\infty^+}) = (1 + 2 \spin , 0)$, representing the asymptotic behaviour of one of the homogeneous solutions. The coefficients $( \tilde \eta_{\infty^-} , \tilde \varsigma_{\infty^-})$, associated with the second linearly independent homogeneous solution, encode information about $\scri^-$, including the peeling decay $\Psi_0 \sim r^{-1}$ and $\Psi_4 \sim r^{-5}$. However, the asymptotic behaviour also includes additional oscillatory terms, as the $\t = \text{constant}$ hypersurfaces in the $(-++)$ configuration (outgoing Kerr coordinates) do not properly foliate $\scri^-$.

The same reasoning should hold for the configuration $(+--)$, but with the roles of $\scri^+$ and $\scri^-$ inverted. Because $\t =$ constant in this configuration corresponds to ingoing Kerr coordinates, the coefficients $( \tilde \eta_{\infty^-} , \tilde \varsigma_{\infty^-}) = (1- 2 \spin , 0)$ yields a homogeneous solution with no oscillatory behaviour toward $\scri^-$, and the oscillations are shifted to the solution behaving as $( \tilde \eta_{\infty^+} , \tilde \varsigma_{\infty^+})$ towards $\scri^+$. The peeling properties for $\Psi_0$ and $\Psi_4$, however, do not show up explicitly as the identification of the Teukolsky master function with the Weyl scalar is geometrically constructed by the particular choice of null tetrads, which in this case, differs from the original Kinnersley tetrads. This fact confirms, once again, a central message of this work: the interpretation of the radial Teukolsky equation in terms of confluent Heun functions should not be de-attached from the underlying geometrical elements underlying the formalism, i.e, the choice of null tetrads and the underlying time foliation upon with the radial functions has their domain of validity.

\subsubsection{Configuration $(---)$ and $(+++)$}\label{sec:--- / +++ NewNot}
This configuration provides a minimalistic change with respect to the outgoing and ingoing Kerr coordinates: it just flips the sign of $a_\infty$, i.e. it defines the coordinate transformation \eqref{eq:GenRadialTrasfo} in terms of a function $h(r)$ which differs from the tortoise coordinate $r^*(r)$ just on the sign of the term going as $r^*(r) \sim r$. 

Intuitively, one might expect that this change should not affect the behaviour at horizon, but only the asymptotic structure. This is indeed observed in the right top panel of Figure~\ref{fig:penrose_diagrams},  where the slices for the configuration $(---)$ (blue) and $(+++)$ (red) differ from the previous case just by their end point at $i^0$, as opposed to $\scri^\pm$. 

This property is confirmed by the asymptotic behaviour of the conformal norm and the extrinsic mean curvature. Both configurations have a space-like surface $(\tilde N < 0)$ asymptotically, with the extrinsic curvature going to zero as $K \sim \mp 1/\sqrt{r}$ for either case. Note that the sign of $K$ (as it goes to zero asymptotically) coincides with the sign of $\nu_\pm$ in the triad $(\mu_c, \mu_h, \nu)$ labelling these surfaces.

Even though these configurations approach $i^0$ as $r\rightarrow \infty$, it is interesting to notice that the asymptotic behaviour derived from the characteristic exponents $(\eta_{\infty^\pm} \varsigma_{\infty^\pm})$ differ significantly from the corresponding values derived in terms of the Boyer-Lindquist coordinate, cf.~eq.~\eqref{eq:BL_coeff_infty}. Even though $\eta^\pm$ has the same structure as in outgoing/ingoing Kerr coordinates, the exponents $\varsigma_{\infty^\pm}$ make the oscillatory behaviour unavoidable as $r \rightarrow \infty$.

\subsubsection{Configuration $(-++)$ and $(+--)$}\label{sec:-++ / +-- NewNot}
Still having the outgoing/ingoing Kerr coordinate as a reference, the configuration $(-++)$ and $(+--)$ flips the sign of the coefficient $b_h$ and $c_h$ with respect to $(--+)$ and $(++-)$. Given the constraint \eqref{eq:constraint_coeff}, these changes also imply a modification in the values for $b_\infty$ and $c_\infty$. Curiously, however, these changes affect the slices only in the interior of the white and black hole regions. For the configuration $(-++)$ the surface $r=\rc$ along $\t = $ constant reaches the Cauchy horizon ${\cal C}_{\rm f}^-$, as opposed to  ${\cal C}_{\rm f}^+$ from the cases \ref{sec:--+ / ++- NewNot} and \ref{sec:--- / +++ NewNot}. In other words, here the Cauchy horizon ${\cal C}_{\rm f}$ is generated by the null vector $\hat \ell^a_-$, whereas the previous examples had $\hat \ell^a_+$ as a null generator. The configuration $(+--)$ follows the same reasoning, with the surface $r=\rc$ being reached at ${\cal C}_{\rm p}^+$ ($\hat \ell^a_+$ as null generator), as opposed to ${\cal C}_{\rm p}^-$ ($\hat \ell^a_-$ as null generator) in the cases \ref{sec:--+ / ++- NewNot} and \ref{sec:--- / +++ NewNot}. These slices are shown in the bottom left panel of Figure~\ref{fig:penrose_diagrams}.

The asymptotic end, however, is still $i^0$ as in \ref{sec:--- / +++ NewNot}. The surface $\t=$ constant for the configurations $(-++)$ and $(+--)$ are asymptotically spacelike, with $\tilde N < 0$ as $r\rightarrow \infty$, and the extrinsic mean curvature behaves as $K \sim \pm 1/\sqrt{r}$. Here, we also observe that the asymptotic sign of $K$ coincides with the one from $\nu_\pm$ in the triad $(\mu_c, \mu_h, \nu)$ labelling the surface. And as in the previous section, the characteristic exponents $(\eta_{\infty^\pm} \varsigma_{\infty^\pm})$ capturing the field's behaviour toward $i^0$ is different from the Boyer-Lindquist case. Therefore, we encounter 3 different sets of configuration --- Boyer-Lindquist, $(---)/(+++)$ and $(-++)/(+--)$ --- where the time slicing approaches $i^0$ as $r\rightarrow \infty$, but each of them with a different asymptotic behaviour. This result suggests that the asymptotic behaviour depends on the particular rate at which the time slices $\t= \text{constant}$ accumulate at $i^0$, and it may allow the use of tools provided by the conformal representation of the cylinder at spatial infinity to study the asymptotic structure of the Kerr spacetime \cite{Hennig:2020rns, Friedrich1987, Friedrich1998, Friedrich2002}.

\subsubsection{Configuration $(-+-)$ and $(+-+)$}\label{sec:-+- / +-+ NewNot}
This final configuration introduces a modification analogous to the one between cases \ref{sec:--+ / ++- NewNot} and \ref{sec:--- / +++ NewNot}. It only flips the sign of the coefficient $a_\infty$ with respect to the previous configuration $(-++)$ and $(+--)$. In a similar way as before, the effect of this modification is to change the hypersurface's asymptotic behaviour. Very interestingly, surfaces $\t = \text{constant}$ for the configuration $(-+-)$ and $(+-+)$ are asymptotically {\em timelike}, with $\tilde N > 0$ as $r\rightarrow \infty$. For the case $(-+-)$, the extrinsic mean curvature vanishes as $K \sim -1/\sqrt{r}$, indicating that the surfaces have past timelike infinity $i^-$ as an asymptotic end. For the case $(+-+)$, on the other hand, $K \sim 1/\sqrt{r}$ and the surfaces approach $i^+$ asymptotically. Therefore, as $r\rightarrow \infty$, the characteristic exponents $(\eta_{\infty^\pm}, \varsigma_{\infty^\pm})$ provide an asymptotic expansion suitable for $i^\pm$. This behaviour is captured in the Carter-Penrose diagram, Figure~\ref{fig:penrose_diagrams}'s bottom right panel.

\section{Comparing the confluent Heun's solution against hyperboloidal}\label{sec:hyp}
The previous section studies the spacetime interpretation for all possible configuration leading the radial Teukolsky equation into the canonical CHE. As observed, even though all Heun slices are horizon-penetrating, none of them is hyperboloidal. In this section, we review the hyperboloidal formalism for the Kerr spacetime with focus on the representation of the solutions in terms of the Heun functions.

As in eq.~\eqref{eq:GenCoordSys}, the transformation from the original Boyer-Lindquist coordinates into hyperboloidal coordinates $\bar{x}^a=( \bar{t}, \sigma,\theta,\bar\phi)$ follows from the height function technique~\cite{Zenginoglu:2007jw}
\bea
\label{eq:Coords_Hyp}
t = \lambda \bigg( \bar{t} - H(\sigma,\theta) \bigg), \quad r = \lambda \dfrac{\rho(\sigma)}{\sigma}, \quad  \phi = \bar \phi - \bar\chi(\sigma),
\eea
with $\lambda$ a length scale of the spacetime, and $H(\sigma,\theta)$ a height function enabling the surfaces $\bar{t}=$ constant to remain spacelike while penetrating the future black hole horizon, and extending towards future null infinity.  
Eq.~\eqref{eq:Coords_Hyp} also introduces a radial compactification, which is absent in the previous sections, allowing us to place future null infinity at the finite value $\sigma = 0$. Besides, it also maps the surfaces $r=\{ 0, \rc, \rh \}$ into $\sigma = \{\sigma_{\rm sing}, \sigma_{\rm c}, \sigma_{\rm h} \}$. 
Finally, the azimutal angular coordinate transforms as in the case of the ingoing Kerr coordinates \eqref{eq:InKerr} with $\bar\chi(\sigma) = \chi(r(\sigma))$, as this choice ensures the regularity of the null vectors at the future black hole horizon. 

The generic form of the line element \eqref{eq:Kerr_BL} in terms of the coordinates $\bar{x}^a=( \bar{t}, \sigma,\theta,\bar\phi)$ is not relevant for the discussion, but an important consequence of the  coordinate transformation \eqref{eq:Coords_Hyp} is that it naturally yields a conformal re-scaling of the line element
\beq
\label{eq:ConfFact}
\d\bar s^2 = \Omega(\sigma)^2 \d s^2,  \quad \Omega(\sigma) = \sigma/\lambda,
\eeq 
with the conformal metric $\d\bar s^2$ regular in the domain $\sigma \in [0,\sigma_{\rm sing} )$. 

A null tetrad basis  $(\bar \ell_+^a, \bar \ell_-^a, \bar m^a, \bar m^\star{}^a)$ associated with the conformal metric $\bar g^{ab} = -2 \bar \ell_+^{\,(a} \bar \ell^{\,b)}_-  + 2 \bar m^{\,(a} \bar m^\star{}^{\,b)}$ is obtained from the original Kinnerlsy tetrad basis via
\bea
\label{eq:conf_tetrad}
\bar \ell^a_+ = \dfrac{\zeta(\sigma)}{\Omega(\sigma)} \ell^a_+, \quad \bar \ell^a_- =  \dfrac{1}{\zeta(\sigma) \Omega(\sigma)} \ell^a_-, \quad \bar m^a = \dfrac{1}{\Omega(\sigma)}  m^a, \quad \bar m^\star{}^a = \dfrac{1}{\Omega(\sigma)} m^\star{}^a.
\eea
We emphasise that $(\bar \ell_+^a, \bar \ell_-^a, \bar m^a, \bar m^\star{}^a)$ is a tetrad basis {\em with respect to the conformal spacetime} $d\bar s^2$, i.e., it arises from a conformal re-scaling of a physical tetrad, together with a boost akin to eq.~\eqref{eq:GenTetrad}, with a boost factor\footnote{To simplify the notation, we omit the explicit reference to radial dependence in some functions, for instance, in the conformal factor $\Omega(\sigma)$ or in the metric function $\Delta(r(\sigma))$. }
\beq
\label{eq:boost_hyp}
\zeta = \dfrac{\Omega\, \Delta}{\lambda}.
\eeq
Eqs.~\eqref{eq:conf_tetrad} and \eqref{eq:boost_hyp} ensure a conformal null tetrad $(\bar \ell_+^a, \bar \ell_-^a, \bar m^a, \bar m^\star{}^a)$ regular in the domain $\sigma \in [0,\sigma_{\rm sing} )$, with the constant $\lambda$ responsible for making the conformal null vectors $(\bar \ell_+^a, \bar \ell_-^a)$ dimensionless.

In the hyperboloidal coordinates $\spinPsi{\spin}(x^a)$, the original Teukolsky master function, is usually transformed into a function $\spinHyperPsi{\spin}(\bar x^a)$ regular for $\sigma \in [0,\sigma_{\rm sing} )$ via~\cite{Zenginoglu:2011jz,PanossoMacedo:2019npm}
\beq
\label{eq:TeukFunc_BL_HypRadial}
\spinHyperPsi{\spin}(\bar x^a) = \Omega \Delta^{-\spin} \,\, \spinPsi{\spin}(x^a),
\eeq
where the transformation $x^a(\bar x^a)$ between Boyer-Lindquist and hyperboloidal coordinates is understood. As in the case of Ingoing Kerr coordinates, eq.~\eqref{eq:TeukFunc_BL_HypRadial} acquires a term $\propto \Delta^{-\spin}$ as a consequence of the boost transformation \eqref{eq:boost_hyp} ensuring regularity at the horizons. The extra factor $\Omega$ results from constructing the master function directly from the conformal spacetime $\bar g_{ab}$ \cite{Gasperin2025}. For instance when $\spin =0$, the term $\Omega \sim 1/r$ is the factor typically introduced to take into account the asymptotic decay of the scalar field, see appendix in \cite{Wald:1984rg} for a discussion in terms of conformal transformations.

A decomposition into the frequency domain directly in hyperboloidal coordinates $\bar x^a$
\beq
\label{eq:Fourier_HyperRadial}
\spinHyperPsi{\spin}(\bar t, \sigma, \bar\theta, \bar \phi)= \dfrac{1}{2 \pi} \int_{-\infty }^{\infty } \sum_{\ell = |\spin|}^{\infty}\sum_{m=-\ell}^{\ell} e^{-\i \omega \bar t} e^{\i m \bar \phi} \spinHyperR{\omega} {\sigma}\spinS{\omega} {\rm d} \omega,
\eeq
yields a conformal radial function $\spinHyperR{\omega} {\sigma}$ relating to the original $\spinR{\omega} {r}$ by
\beq
\label{eq:HyperRadialtransf}
\spinR{\omega} {r(\sigma)} = \bar{\cal Z}(\sigma)\, \spinHyperR{\omega} {\sigma}, \quad
\bar {\cal Z} = \Omega \Delta^{-\spin} \, e^{-\i \omega \lambda H} e^{\i m \bar\chi}.
\eeq

\subsection{The minimal gauge}
The height function $H(\sigma,\theta)$ and the radial function $\rho(\sigma)$ encode the degrees of freedom associated with the hyperboloidal transformation \eqref{eq:Coords_Hyp}. We fix these functions within the so-called minimal gauge class \cite{PanossoMacedo:2023qzp,PanossoMacedo:2019npm}. For that purpose we first consider the radial degree of freedom, where the radial function $\rho(\sigma)$ assumes the form
\beq
\label{eq:rho}
\rho(\sigma) = \rho_0 + \rho_1\sigma, \quad \rho_0 = \dfrac{\rh}{\lambda} -\rho_1.
\eeq
Eq.~\eqref{eq:rho} arises from imposing the most simple form for the differential $dr = r'(\sigma) d\sigma$, with $r'(\sigma) =  - \lambda \rho_0 \sigma^{-2}$, whereas the particular choice for $\rho_0$ ensures that the radial transformation in \eqref{eq:Coords_Hyp} maps the black hole horizon $\rh$ into sigma $\sigma_{\rm h} = 1$. The Cauchy horizon and singularity then assume the values
\beq
\sigma_{\rm c}= \dfrac{1 - \lambda \rho_1/\rh}{ \kappa^2 - \lambda \rho_1/\rh}, \quad \sigma_{\rm sing}= 1 - \dfrac{\rh}{\lambda \rho_1}.
\eeq
From the radial transformation in eq.~\eqref{eq:Coords_Hyp}, we also introduce the dimensionless tortoise coordinates
\bea
x(\sigma) &=& \dfrac{r^*(r(\sigma))}{\lambda} \nn \\
\label{eq:x}
&=& x_{\infty}(\sigma) + x_{\rm h}(\sigma) + x_{\rm c}(\sigma) + x_{\rm cont},
\eea
with $x_{\infty}(\sigma)$, $x_{\rm h}(\sigma)$ and $x_{\rm c}(\sigma)$ terms which are singular, respectively, at future null infinity $\sigma = 0$, the black hole horizon $\sigma = 1$ and the Cauchy horizon $\sigma=\sigma_c$, and given by \cite{PanossoMacedo:2023qzp}
\bea
\label{eq:x_terms}
x_{\infty}(\sigma) = \dfrac{\rho_0}{\sigma} - \dfrac{2M}{\lambda} \ln \sigma, \quad
x_{\rm h}(\sigma) = \dfrac{1}{2 \lambda \varkappa_h } \ln(1- \sigma), \quad x_{\rm c}(\sigma) = \dfrac{1}{2 \lambda \varkappa_c } \ln\left(1- \dfrac{\sigma}{\sigma_c}\right).
\eea
We emphasise that, even though the entire expression for $x(\sigma)$ is derived directly from $r^*(r)$, the individual terms in eq.~\eqref{eq:x_terms} do not relate necessarily to the respective terms in $r^*(r)$. Indeed, from eq.~\eqref{eq:tort_r} one would obtain
\bea
\label{eq:r* infty}
\dfrac{r^*_\infty(r(\sigma))}{\lambda} &=& x_{\infty}(\sigma) +  \dfrac{2 M}{\lambda} \ln \left( 1- \dfrac{\lambda \rho_1}{\rh} (1- \sigma) \right) + \rho_1, \\
\dfrac{r^*_h(r(\sigma))}{\lambda}&=& x_{\rm h}(\sigma) - \dfrac{1}{2\lambda \varkappa_h}\ln \left(  1- \dfrac{\lambda \rho_1}{\rh} (1- \sigma) \right) + \dfrac{1}{2\lambda \varkappa_h} \ln\left(  1 - \dfrac{\lambda \rho_1}{\rh} \right), \\
\label{eq:r* c}
\dfrac{r^*_c(r(\sigma))}{\lambda}&=& x_{\rm c}(\sigma) - \dfrac{1}{2\lambda \varkappa_c}\ln \left(  1- \dfrac{\lambda \rho_1}{\rh} (1- \sigma) \right) + \dfrac{1}{2\lambda \varkappa_c} \ln\left(  1 - \dfrac{\lambda \rho_1}{\rh} \right).
\eea
Even though each individual singular term in $r_*(r(\sigma))$ and $x(\sigma)$ may not agree (unless $\rho_1 = 0$), the terms with $\ln \left( 1- \dfrac{\lambda \rho_1}{\rh} (1- \sigma) \right)$ cancel out exactly upon summing eqs.~\eqref{eq:r* infty}-\eqref{eq:r* c}, and one identifies the constant term
\beq
x_{\rm cont} = \rho_1 + \left( \dfrac{1}{2\lambda \varkappa_h} + \dfrac{1}{2\lambda \varkappa_c} \right) \ln\left(  1 - \dfrac{\lambda \rho_1}{\rh} \right).
\eeq
A similar cancelation occurs in the azimutal function $\bar \chi(\sigma) = \chi(r(\sigma))$, which then yields
\beq
\bar \chi(\sigma) = \dfrac{\Omega_h}{2 \varkappa_h } \ln(1- \sigma) + \dfrac{\Omega_c}{2  \varkappa_c } \ln\left(1- \dfrac{\sigma}{\sigma_c}\right).
\eeq

Turning our attention to the time transformation in eq.~\eqref{eq:Coords_Hyp}, the height function in the minimal gauge follows directly from the terms in eq.~\eqref{eq:x_terms} by a simple change in the sign in the $x_{\infty}$ term~\cite{PanossoMacedo:2023qzp}, i.e.,
\beq
\label{eq:Height_MinGauge}
H(\sigma, \theta) = -x_{\infty}(\sigma) + x_{\rm h}(\sigma) + x_{\rm c}(\sigma).
\eeq
With the above considerations, there remains only the parameter $\rho_1$ to fully fix the hyperboloidal foliation. Two choices for $\rho_1$ are of relevance. The most simple option is just to set $\rho_1 = 0$, but we can also use it to fix further relevant surfaces, such as the Cauchy horizon, to a constant coordinate value. We consider the former in sec.~\ref{sec:radial fix} and the latter in sec.~\ref{sec:cauchy fix}


\subsubsection{Radial fixing}\label{sec:radial fix}
\begin{figure*}[t] 
\includegraphics[width=0.5\textwidth]{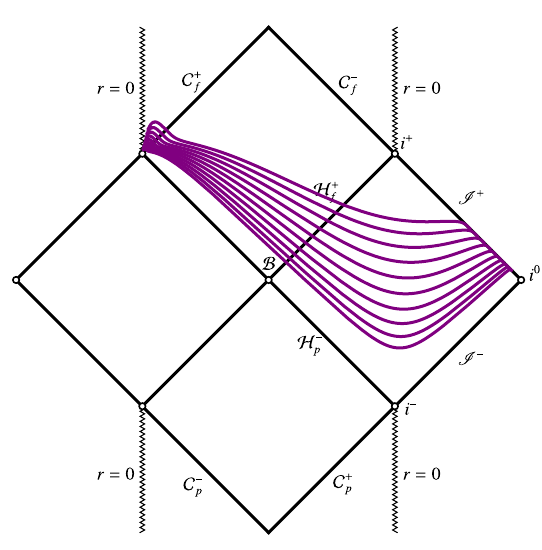}
\includegraphics[width=0.5\textwidth]{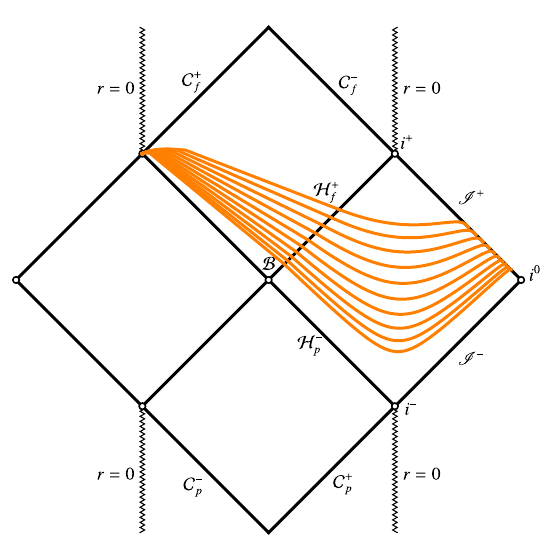}
  \caption{Carter-Penrose diagrams representing hyperboloidal slices in the minimal gauge class. {\em Left Panel:} In the radial fixing gauge, the slices not only intersect future null infinity $\scri^+$ and the black hole horizon ${\cal H}_{\rm f}^+$, but the Cauchy horizon ${\cal C}_{\rm f}^+$ is also a regular surface, with its coordinate value $\sigma_{\rm c} = \kappa^{-2}$ depending on the black hole rotation parameter. Across ${\cal C}_{\rm f}^+$, the hypersurfaces change character as and they all accumulate into singular point as  $\sigma \rightarrow \infty$ $(r\rightarrow 0)$. {\em Right Panel:} In the Cauchy fixing gauge, the Cauchy horizon is fixed at $\sigma \rightarrow \infty$ and all the slices accumulate into the singular point without crossing ${\cal C}_{\rm f}^+$. This accumulation into a singular point as $\sigma \rightarrow \infty$ in both cases leads to a violation of conditions \eqref{eq:cond_Allternative_Heun} in the confluent Heun equation.
}
    \label{fig:penrose_diagram_hyp}
\end{figure*}

The radial fixing gauge is characterised by setting
\bea
\rho_1 =0 \Longrightarrow \rho(\sigma) = \dfrac{\rh}{\lambda} \quad ({\rm constant}),
\eea
with the left panel of Fig.~\ref{fig:penrose_diagram_hyp} showing examples of the corresponding hyperboloidal time slices.

This choice corresponds to a direct compactification of the radial coordinate
\bea
r(\sigma)= \dfrac{r_h}{\sigma}.
\eea
In particular, the coordinate location of the Cauchy horizon in the compact coordinate depends on the spin parameters $\sigma_{\rm c} = \kappa^{-2}$, whereas the surface $r=0$ is pushed to $\sigma_{\rm sing} \rightarrow \infty$. 
Some geometrical properties along these hypersurfaces are read from the conformal null vectors \eqref{eq:conf_tetrad}
\bea
\bar \ell^a_+ &=& \dfrac{2 \rh ^2}{\lambda^2} (1 + \kappa^2)  \bigg(1 + \kappa^2 (1 - \sigma)  \bigg) \delta^a_{\bar t} - \dfrac{r_h}{\lambda} (1 - 
   \sigma) (1 - \kappa^2 \sigma)\delta^a_\sigma + \dfrac{2 r_h \kappa}{\lambda}  \delta^a_{\bar\phi}, \\
 \bar \ell^a_- &=&\dfrac{\lambda^2}{\rh^2 \left(1 +  \kappa^2 \sigma^2 \cos^2\theta\right)} \left(  \left(1 + \sigma + \kappa^2 \sigma \right) \delta^a_{\bar t} +\dfrac{\lambda \sigma^2}{2 r_h} \delta^a_\sigma \right).
\eea
Specifically, not only does one confirm their regularity at $\sigma = 0$
\bea
\left. \bar \ell^a_+ \right |_{\sigma=0} = \dfrac{2 \rh^2}{\lambda^2 } (1 + \kappa^2)^2\delta^a_{\bar t} - \dfrac{\rh}{\lambda} \delta^a_\sigma +  \dfrac{2 r_h \kappa}{\lambda} \delta^a_{\bar\phi},\quad
\left.  \bar \ell^a_- \right|_{\sigma=0} = \dfrac{\lambda^2}{\rh^2}\delta^a_{\bar t},
\eea
but also that this surface is generated by the ingoing null vector $\bar \ell^a_- $, which confirms that $\sigma = 0$ corresponds to $\scri^+$ as expected. Similarly, the conformal null vectors are regular at $\sigma = 1$  and $\sigma= \kappa^{-2}$
\bea
&\left. \bar \ell^a_+ \right|_{\sigma=1} = \dfrac{2 r_h^2 }{\lambda^2}   (1 + \kappa^2)\delta^a_{\bar t} +  \dfrac{2 r_h \kappa}{\lambda} \delta^a_{\bar\phi} ,\quad
\left. \bar \ell^a_- \right|_{\sigma=1} =  \dfrac{\lambda^2}{\rh^2 \left(1+\kappa^2 \cos \theta^2\right)}\bigg{(}(2+\kappa^2)\delta^a_{\bar t} +\dfrac{\lambda}{2 \rh}\delta^a_\sigma \bigg{)}, \\
&\left. \bar \ell^a_+\right|_{\sigma = \kappa^{-2}} = \dfrac{2 \rh^2}{\lambda^2} \kappa^2(1 + \kappa^2)\delta^a_{\bar t} +  \dfrac{2 r_h \kappa}{\lambda} \delta^a_{\bar\phi},\;
\left. \bar \ell^a_-\right|_{\sigma = \kappa^{-2}} =\dfrac{\lambda^2}{\rh^2 \left(\kappa^2+ \cos \theta^2\right)}\bigg{(}(1 + 2\kappa^2)\delta^a_{\bar t} +\dfrac{\lambda}{2 \rh \kappa^2}\delta^a_\sigma \bigg{)},
\eea
with $\bar \ell^a_+$ generator of the surfaces in both cases, confirming that $\sigma = 1$  and $\sigma= \kappa^{-2}$ correspond, respectively, to the black-hole horizon ${\cal H}_{\rm f}^+$ and the Cauchy horizon ${\cal C}_{\rm f}^+$.

It is interesting to observe that the radial Teukolsky equation for $\spinHyperR{\omega} {\sigma}$ assumes the alternative form for the Heun function \eqref{eq:CHE_compact}. However, the corresponding coefficients
\bea
&&A{}_{0}^{(\sigma)}=1-\mathfrak{s}- \dfrac{\i (m \Omega_h - \omega )}{\varkappa_h}, \qquad A{}_{1}^{(\sigma)}=1-\mathfrak{s}- \dfrac{\i (m \Omega_c -\omega )}{\varkappa_c}, \nn \\
&&E{}_0{}^{(\sigma)}= 2 (1 + \mathfrak{s}- \i r_h (1 + \kappa^2) \omega ), \qquad  E{}_1{}^{(\sigma)}= -2 \i r_h \omega, \nn \\
&&C{}_0{}^{(\sigma)}=-  \dfrac{\kappa^2(1+(1- \i \omega)\mathfrak{s}+r_h\kappa^2 \Omega_h  Alm)   }{2 r_h \varkappa_h}+ \dfrac{
  \i r_h \kappa \Omega_h (-1 + 2 \kappa^4) \omega}{\varkappa_h} + \nn \\
&& \hspace{1.3cm} + \dfrac{
 r_h^2 \kappa^3 \Omega_h (4 + \kappa^2(11 + 
    8 \kappa^2)) \omega^2}{\varkappa_h} - \dfrac{
 m \kappa^2 \Omega_h( \i +r_h (2+3 \kappa^2)\omega) }{\varkappa_h} , \nn \\
&&C{}_1{}^{(\sigma)}=1- \dfrac{Alm}{-1 + \kappa^2}-\dfrac{2 \i  r_h (-2 + \kappa^4) \omega}{-1 + \kappa^2} +\dfrac{r_h^2 (8 + \kappa^2 (11+ 
    4 \kappa^2)) \omega^2}{-1 + \kappa^2} +  \\
&& \hspace{1.3cm} -\dfrac{m  \Omega_c (\i + r_h(3+2\kappa^2)\omega)}{\varkappa_c}  + \dfrac{\mathfrak{s}(1+ \i \kappa^2 \omega)}{2 r_h \kappa^2 \varkappa_c}, \nn \\
&&D{}_0{}^{(\sigma)}= -1 - 2 \i m \kappa - \kappa^2 - Alm (1 + \kappa^2) + 
  4 \i r_h \omega+ r_h \omega(2 \i \kappa^2 (3 + 2 \kappa^2) - 6 m (\kappa + \kappa^3) + \nn \\
&& \hspace{1.3cm}+ r_h (1 + \kappa^2) (8 + 11 \kappa^2 +   8 \kappa^4) \omega)  - \mathfrak{s} - \kappa^2 \mathfrak{s}, \nn \\
&&D{}_1{}^{(\sigma)}= -Alm + r_h \omega (-2 m \kappa + r_h (4 + 7 \kappa^2 + 4 \kappa^4) \omega - 
     2 \i (1 + \kappa^2) \mathfrak{s}), \nn \\
&&B{}_0{}^{(\sigma)}=0, \qquad B{}_1{}^{(\sigma)}=0, \qquad D{}_2{}^{(\sigma)}=0, \qquad  D{}_3{}^{(\sigma)}= 0 \nn
\eea
 {\em do not} satisfy the conditions \eqref{eq:cond_Allternative_Heun} ensuring that the point $\sigma_{\rm sing} \rightarrow \infty$ is a regular singular point. 

An alternative way to observe the non-Heun form for the radial equation along a hyperboloidal slice is to re-express it in terms of the coordinate $z$ in eq.~\eqref{eq:z_of_r_BH} via
\beq
z(\sigma) = \dfrac{1}{\sigma} \dfrac{1-\kappa^2 \sigma}{1- \kappa^2} \longleftrightarrow \sigma(z) = \dfrac{1}{z + \kappa^2 (1-z)}.
\eeq
The resulting equation {\em does not} assume the canonical form \eqref{eq:canonoical_CHE_intro}. 

To understand the differences between hyperboloidal slices and Heun slices, we can compare the expressions between height functions in the hyperboloidal case \eqref{eq:Height_MinGauge} against eq.~\eqref{eq:Height_HeunSlices} leading the Heun slices.
In terms of the notation introduce in Table \ref{tab:HeunCoeffSpacetime}, the radial fix minimal gauge has coefficients $(a_\infty, b_\infty, b_h, b_c) = (-1, -1, +1, +1)$, $(c_h, c_c) = ( +1, +1)$  and $(d_h, d_c) = ( +1, +1)$, which is almost the same as the configuration $(+++)$. The only difference lies in the coefficient $b_\infty$, required to ensure the surface $\bar{t} =$ constant asymptotically reaches $\scri^+$ as opposed to $i^0$.

Alternatively, we can express the map \eqref{eq:HyperRadialtransf} between the hyperboloidal and Boyer-Lindquist radial function in terms of
\beq
\label{eq:Z_Hyp}
\bar {\cal Z}(\sigma(r) ) \propto r^{\bar \mu_{\infty}} \left( r -\rh \right)^{\bar \mu_h}  \left( r -\rc \right)^{\bar \mu_c} e^{\bar \nu r}
\eeq
with
\beq
\label{eq:Z_coeff_Radial}
\bar \mu_{\infty} = -1 + 4 \i M \omega, \quad \bar \mu_h = - \spin - \dfrac{\i \left( \omega - m \Omega_h \right)}{2 \varkappa_h} , \quad \bar \mu_c =  - \spin - \dfrac{\i \left( \omega - m \Omega_c \right)}{2 \varkappa_c}  , \quad \bar \nu = \i \omega
\eeq
As anticipated, the triad $(\bar \mu_c, \bar \mu_h, \bar \nu)$ directly relates to the configuration $(+,+,+)$. However, the hyperboloidal mapping brings further contribution with $\bar \mu_\infty \neq 0$, absent in the case of the s-homotopic transformation \eqref{HomTrasfR}. This extra term $\propto r^{\bar \mu_\infty}$ --- crucial to ensure the desired asymptotic behaviour for the hyperboloidal slices --- also affects the behaviour of the time slices around $r\rightarrow 0$, making this limit an irregular singular point in the radial equation. 

This property is confirmed geometrically by observing the behaviour of the hypersurfaces $\bar t =$ constant in the left panel of Fig.~\ref{fig:penrose_diagram_hyp} in the region beyond the Cauchy horizon. The hypersurfaces change character and, as $\sigma \rightarrow \infty$, they all accumulate in the irregular point connecting $r=0$ with $i^+$.

While the solutions to the hyperboloidal radial equation are not Heun functions, the latter can still be used to deduce properties of the former. For instance, the characteristic exponents for the hyperboloidal radial functions $\spinHyperR{\omega} {\sigma}$ are derived from eq.~\eqref{eq:HyperRadialtransf} as
\bea
\label{eq:Hyp+CharcExp}
&\bar \varrho_{{\cal H}_{\rm p/f}}& = \varrho_{{\cal H}_{\rm p/f}} - \bar \mu_h = 
\left\{
\begin{array}{cc}
0 & ({\cal H}_{\rm f}) \\
\spin +  \dfrac{\i \left( \omega - m \Omega_h\right)}{\varkappa_h} & ({\cal H}_{\rm p}) \\
\end{array}
\right. \\
& \bar \eta_{\infty^{\pm }}& = \eta_{\infty^{\pm}} + \bar \mu_\infty + \bar \mu_h  + \bar \mu_c=  
\left\{
\begin{array}{cc}
0 & (\scri^+) \\
-2 \spin + 4 \i M \omega  & (\scri^-) \\
\end{array}
\right. \\
& \bar \varsigma_{\infty^\pm}& = \varsigma_{\infty^\pm}  - \nu= \left\{
\begin{array}{cc}
0 & (\scri^+) \\
- 2 \i \omega  & (\scri^-) \\
\end{array}
\right. .
\eea
The vanishing of all characteristic exponents at $\scri^+$ and ${\cal H}_{\rm f}^+$ is a direct consequence of the hyperboloidal foliation, which is naturally adapted to these surfaces, but oscillations are expected at $\scri^-$ and ${\cal H}_{\rm p}^-$.

\subsubsection{Cauchy fixing}\label{sec:cauchy fix}
In the radial fixing gauge, the coordinate location of Cauchy horizon $\sigma_{\rm c}$ depends explicitly on the black hole rotation parameter $\kappa$. A second possibility within the minimal gauge class is to exploit the free parameter $\rho_1$ in the radial compactification \eqref{eq:rho} to set the Cauchy horizon at a fixed coordinate value independent of $\kappa$. Thus, the so-called Cauchy fixing gauge \cite{PanossoMacedo:2019npm} is characterised by setting
\bea
\rho_1 =\dfrac{\rc}{\lambda} \Longrightarrow \rho(\sigma) = \dfrac{\rh-\rc}{\lambda} + \dfrac{\rc \sigma}{\lambda}.
\eea
From the above form of $\rho(\sigma)$, it follows the radial transformation
\bea
\label{eq:r_of_sigma_Cauchy}
r(\sigma)= \dfrac{\rh-\rc}{\sigma} +\rc \Longleftrightarrow \sigma (r)= \dfrac{\rh-\rc}{r-\rc},
\eea
which is equivalent to a direct compactification of the coordinate \eqref{eq:z_of_r_BH} as $z=1/\sigma$, mapping $r=\rc$ into $\sigma_{\rm c}\rightarrow \infty$. The surface $r=0$, on the other hand, is mapped into $\sigma_{\rm sing} =1- \kappa^{-2}$, i.e.  $\sigma_{\rm sing}<0$ lies outside the domain $\sigma \in [0, \infty)$ connecting $\scri^+$ to the Cauchy horizon.

The right panel of Fig.~\ref{fig:penrose_diagram_hyp} shows examples of the corresponding hyperboloidal time slices. As opposed to the radial fixing gauge, the surfaces do not cross over the Cauchy horizon, and they accumulate directly in the singular point connecting ${\cal C}_{f}^+$ and $i^+$. As in the previous section, the conformal null vectors \eqref{eq:conf_tetrad} 
\bea
\bar \ell^a_+ &=&  \dfrac{2\rh^2}{\lambda^2} (1 + \kappa^2)\delta^a_{\bar t}- \dfrac{\rh}{\lambda} (1 -  \sigma) (1 - \kappa^2 )\delta^a_\sigma + \dfrac{2\rh \kappa}{\lambda}  \delta^a_{\bar\phi}, \\
\bar \ell^a_- &=& \dfrac{\lambda^2}{ \rh^2 (1-\kappa^2) \Bigg( \bigg( 1 - \kappa^2  \left(1-\sigma\right) \bigg)^2 + \kappa^2 \sigma^2 \cos^2\theta \Bigg)} \Bigg(  \bigg( 1+\sigma - \kappa^2 \left(1-\sigma\right) \bigg) \delta^a_{\bar t} + \dfrac{\lambda \sigma^2}{2\rh} \delta^a_\sigma\Bigg)
%
\eea
contain information about some geometrical properties along these hypersurfaces. For instance, they are regular at $\sigma = 0$
\bea
\left. \bar \ell^a_+ \right |_{\sigma=0} = \dfrac{2\rh^2}{\lambda^2} (1 + \kappa^2)\delta^a_{\bar t}- \dfrac{\rh}{\lambda} (1 - \kappa^2 )\delta^a_\sigma + \dfrac{2\rh \kappa}{\lambda}  \delta^a_{\bar\phi},\qquad
\left.  \bar \ell^a_- \right|_{\sigma=0} = \dfrac{\lambda^2}{\rh^2 (1-\kappa^2)^2}\delta^a_{\bar t},
\eea
with the ingoing null vector $\bar \ell^a_- $ generating $\scri^+$, as expected. They are also regular at $\sigma = 1$  
\bea
&\left. \bar \ell^a_+ \right|_{\sigma=1} =\dfrac{2 \rh^2}{\lambda^2} (1 + \kappa^2)\delta^a_{\bar t}+ \dfrac{2 \rh}{\lambda}\kappa \delta^a_{\bar\phi},\quad
\left. \bar \ell^a_- \right|_{\sigma=1} =\dfrac{\lambda^2}{\rh^2 (1-\kappa^2)(1+\kappa^2 \cos \theta^2)}\bigg{(}2\delta^a_{\bar t} +\dfrac{\lambda}{2 \rh}\delta^a_\sigma \bigg{)}, 
\eea
with $\bar \ell^a_+$ generating  ${\cal H}_{\rm f}^+$. However, the conformal null vectors $\bar \ell^a_+$  and  $\bar \ell^a_-$ are not defined as $ \sigma \rightarrow \infty$ confirming the singular character of the surface corresponding to the Cauchy horizon $\mathcal{C}_{\rm f}^+$.

\medskip
When comparing the radial Teukolsky equation for $\spinHyperR{\omega} {\sigma}$ in the Cauchy fixing gauge against the alternative form for the Heun function \eqref{eq:CHE_compact}, one observes the absence of factors $\sim 1/(\sigma - \sigma_0)$ because $\sigma_0 = \sigma_{\rm c}\rightarrow \infty$. The remaining coefficients 
\bea
&A{}_{1}^{(\sigma)}=1-\mathfrak{s}+ \dfrac{\i (m \Omega_h - \omega )}{\varkappa_h},  \quad E{}_0{}^{(\sigma)}=2 (1 - \i r_h (1 + \kappa^2) \omega + \mathfrak{s}), \quad  E{}_1{}^{(\sigma)}= 2 \i r_h (-1 + \kappa^2) \omega, \quad B{}_1{}^{(\sigma)}=0, \nn \\
&C{}_1{}^{(\sigma)}=1 +\mathfrak{s}+ Alm + \dfrac{4 \i r_h \omega}{-1+\kappa^2}+ \dfrac{r_h^2 (8 + \kappa^2(7 + \kappa^2))\omega^2}{-1+ \kappa^2}+\dfrac{ m \Omega_h (\i + r_h (3+ \kappa^2)\omega)}{\varkappa_h}, \nn \\
&D{}_0{}^{(\sigma)}=-1-\mathfrak{s}-Alm - \dfrac{4 \i r_h \omega}{-1+ \kappa^2}-\dfrac{r_h^2( 8 + \kappa^2(7+\kappa^2) )\omega^2}{-1+\kappa^2} -\dfrac{ m \Omega_h (\i + r_h (3+ \kappa^2)\omega)}{\varkappa_h}\nn \\
&D{}_1{}^{(\sigma)}= -Alm + r_h \omega (4 r_h \omega - 
    2 \i \mathfrak{s}+ \kappa (-2 m + 3 r_h \kappa \omega - 
       2 \i \kappa (1 + \mathfrak{s}))), \qquad D{}_2{}^{(\sigma)}=0, \qquad  D{}_3{}^{(\sigma)}= 0, \nn
\eea
and, as in the previous case, they {\em do not} satisfy the conditions \eqref{eq:cond_Allternative_Heun} ensuring that the point $\sigma \rightarrow \infty$ is a regular singular point. As commented, the right panel of fig.~\ref{fig:penrose_diagram_hyp} and the behaviour of the conformal null vectors in this limit confirms the irregular character of this point. 

A systematic comparison of this height functions against eq.~\eqref{eq:Height_HeunSlices} leading the Heun slices, however, requires re-formulating the expression in terms of the non-compact radial $r$. In particular, Eq.~\eqref{eq:r_of_sigma_Cauchy} shows that $\sigma(r)$ and $1-\sigma(r)$ re-introduce terms $\sim 1- \dfrac{\rc}{r}$ and therefore the contributions from the Cauchy horizon coincide with those from future null infinity and the black hole horizon via
\beq
h(r) = \lambda H(\sigma (r)) = - r - 2 M \ln\left(\dfrac{r}{\rh}\right) + \dfrac{1}{2\varkappa_h} \ln\left( 1 - \dfrac{\rh}{r}\right) - \left( 2M + \dfrac{1}{2\varkappa_h}\right) \ln \left( 1- \dfrac{\rc}{r}\right) + H_{\rm const}, 
\eeq  
with $H_{\rm const}$ independent of $r$. In terms of the notation from Table \ref{tab:HeunCoeffSpacetime}, we obtain  $(a_\infty, b_\infty, b_h, b_c) = (-1, -1, 1, -1 + 2\kappa^{-2} )$, $(c_h, c_c) = ( 1, 1)$  and $(d_h, d_c) = ( 1, 1)$.

Similarly, we can express the map \eqref{eq:HyperRadialtransf} as in eq.~\eqref{eq:Z_Hyp} with coefficients
\beq
\label{eq:Z_coeff_Cauchy}
\bar \mu_{\infty} = 0, \quad \bar \mu_h = - \spin - \dfrac{\i \left( \omega - m \Omega_h \right)}{2 \varkappa_h} , \quad \bar \mu_c = -\bar \mu_h -1 + 2 \i  M \omega- 2 \spin  , \quad \bar \nu = \i \omega
\eeq
We recall, that the mapping \eqref{eq:Z_Hyp} with coefficients \eqref{eq:Z_coeff_Cauchy} reproduce exactly the Ansatz introduced by the Leaver's seminal work \cite{Leaver1985}, which allowed the spacetime interpretation of his approach as the frequency domain representation of Teukolsky equation in the hyperboloidal Cauchy fixing minimal gauge \cite{PanossoMacedo:2019npm}.

Since the coefficients at future null infinity and the black hole horizon agree in in the radial and Cauchy fixing gauges --- cf. eqs.~\eqref{eq:Z_coeff_Radial} and \eqref{eq:Z_coeff_Cauchy}, respectively --- the characteristic exponents dictating the behaviour of $\spinHyperR{\omega} {\sigma}$ in the black hole exterior region agree with the values and interpretation from eq.~\eqref{eq:Hyp+CharcExp}.

\subsection{Beyond the minimal gauge}
The arguments in the previous sections are not restricted the particular choice of hyperboloidal foliation in the minimal gauge, but it is a generic feature for any hyperboloidal slice. Indeed, any hyperboloidal foliation follows from a height function \cite{PanossoMacedo:2019npm}
\beq
H(\sigma, \theta) = H_{\rm mg}(\sigma) + A(\sigma, \theta),
\eeq
with $H_{\rm mg}(\sigma)$ the minimal gauge height function from eq.~\eqref{eq:Height_MinGauge} and $A(\sigma, \theta)$ a {\em regular function} along its domain of dependence. Such a change in the time foliation impacts the radial function via a modification in the mapping \eqref{eq:HyperRadialtransf}  
\beq
\bar {\cal Z} \rightarrow \bar {\cal Z} e^{-i \lambda \omega A(\sigma)},
\eeq
where we have restricted ourselves to transformations only along the radial direction. Any attempt to fix $A(\sigma)$ so that the resulting radial equation is in the canonical Heun form will inevitably arises from a singular function $A(\sigma)$, e.g. with terms $\sim \ln(\sigma)$, and which would map the time foliation back into one of the eight configurations discussed in sec.~\ref{sec:Heun Slices}.

\section{Conclusions}\label{sec:conclusion}
In this work, we reviewed the use of Heun functions in describing solutions to the radial Teukolsky equation. Typically, Heun functions are studied through mappings applied solely at the radial coordinate level, using the so-called s-homotopic transformations. By emphasising the geometrical role played by coordinate transformations at the spacetime level, a key outcome of this study is the introduction of the so-called Heun slices.

Specifically, by defining a new coordinate system $\GenCoord$ with appropriate transformations of the time and angular coordinates, as well as a particular choice of null tetrad, we demonstrated that the representation of the radial Teukolsky equation as a confluent Heun equation in the canonical form \eqref{eq:canonoical_CHE_intro} naturally follows from a frequency decomposition with the coordinate $\t$ as a reference. This approach enables us to interpret certain properties of the Heun functions as consequences of the geometrical structure of the hypersurfaces $\t = \text{constant}$, without relying on traditional s-homotopic methods for the radial functions. In particular, we paid special attention to the solutions’ properties near the horizons and at asymptotic distances, connecting the local analytical structure of the Heun functions to the geometry of hypersurfaces $\t = \text{constant}$.

More specifically, there are a total of eight distinct transformations mapping the radial Teukolsky equation in the traditional Boyer–Lindquist coordinates into the canonical confluent Heun equation. When expressed in terms of spacetime hypersurfaces, these transformations correspond to four distinct pairs of time-reversed configurations, which we labelled as: $(--+)/(++-)$; $(---)/(+++)$; $(-++)/(+--)$; and $(-+-)/(+-+)$. Each triplet of $+$ and $-$ signs corresponds to a particular choice of exponents $(\mu_c, \mu_h, \nu)$ responsible for homotopically transforming the radial Teukolsky equation into the canonical confluent Heun form, cf. eq. \eqref{HomTrasfR} and Table \ref{tab:HeunCoeff}. Alternatively, one can interpret these configurations as representing different spacetime foliations and null tetrads, cf. eqs.~\eqref{eq:GenCoordSys}, \eqref{eq:GenTetrad}, \eqref{eq:Height_HeunSlices}–\eqref{eq:boost_zeta}, and Table \ref{tab:HeunCoeffSpacetime}. The most well-known cases are the outgoing $(--+)$ and ingoing $(++-)$ Kerr coordinates.

We observed that all configurations give rise to horizon-penetrating slices, with the signs associated with the exponents $(\mu_c, \mu_h)$ directly related to the null vectors generating the horizons. These are the principal null vectors $\ell^a_+$ and $\ell^a_-$, corresponding to outgoing and ingoing directions, respectively. Thus, the relatively simple behaviour of Heun functions around the regular singular points, expressed by the vanishing characteristic exponents in eq.~\eqref{eq:CharExp_rho}, arises because the Heun slice $\t = \text{constant}$ is naturally adapted to the horizon geometry.

Our analysis also revealed that the asymptotic behaviour of radial perturbations is sensitive to the chosen coordinate slicing. Different slicing configurations produce distinct asymptotic behaviours. For instance, in the cases of outgoing $(--+)$ or ingoing $(++-)$ Kerr coordinates, the asymptotic limit $r \rightarrow \infty$ corresponds to future or past null infinity, respectively, as expected. Here, the signs $+$ and $-$ in the exponent $\nu$ within the triplet are directly associated with the asymptotic geometry along $\t = \text{constant}$ being $\scri^+$ or $\scri^-$, respectively. Because such slices are naturally adapted to foliate null infinity, the corresponding characteristic exponents \eqref{eq:CharExp_sigeta} from the asymptotic expansion of the Heun functions take a regular, non-oscillatory form, consistent with the peeling properties of the Weyl scalars.

Similarly, in configurations $(---)/(+++)$ and $(-++)/(+--)$, $r \rightarrow \infty$ represents spatial infinity. However, each of them predict different characteristic exponents \eqref{eq:CharExp_sigeta} for the asymptotic behaviour of the Heun functions, suggesting that the behaviour around $i^0$ depends on the particular rate at which the time slices $\t = \text{constant}$ accumulate at spatial infinity. A more detailed understanding and formal quantification of these properties will require further study; however, this research direction offers new perspectives on using the conformal representation of the cylinder at spatial infinity in Kerr spacetime. This is particularly relevant given the constraints imposed by current formulations, such as those relying on conformal Gauss geodesics~\cite{Friedrich1987, Friedrich1998, Friedrich2002} or those limited to the massless scalar field~\cite{Hennig:2020rns}.

The Heun slices resulting from the last pair of configurations $(-+-)/(+-+)$ are also noteworthy, as the limit $r \rightarrow \infty$ corresponds to timelike infinity, with the $+$ and $-$ signs in the exponent $\nu$ determining whether it is $i^+$ or $i^-$. As with spatial infinity, it is expected that the behaviour around $i^\pm$ depends on the rate at which the time slices $\t = \text{constant}$ accumulate towards timelike infinity, with the characteristic exponents \eqref{eq:CharExp_sigeta} providing the asymptotic expansion along this Heun slice. As in the case of $i^0$, further formal studies are needed, particularly given the growing interest in developing asymptotic frameworks for gravitational scattering~\cite{Compere:2023qoa}.

In addition to the coordinate transformations associated with Heun solutions, we also explored the use of hyperboloidal coordinates in the study of the Teukolsky equation. Strictly speaking, the radial Teukolsky equation resulting from a hyperboloidal slice does not assume a Heun form. However, when expressed in terms of a compact radial coordinate $\sigma \sim 1/r$, the hyperboloidal radial equation takes a form reminiscent of a Heun equation in the alternative general form \eqref{eq:cond_Allternative_Heun}. Nonetheless, the corresponding coefficients violate the constraints expected for the conformal Heun equation \eqref{eq:CharExp_rho}, ensuring the regularity of the singular point $\sigma \rightarrow \infty$.

Initially, we focused on the relationship between Heun functions and hyperboloidal slices within the minimal gauge class~\cite{PanossoMacedo:2019npm, PanossoMacedo:2023qzp}, which offers two possible strategies for treating the Cauchy horizon. The radial-fixing strategy compactifies the Boyer–Lindquist radial coordinate, making the Cauchy horizon a regular surface within the foliation and mapping the surface $r = 0$ to $\sigma \rightarrow \infty$. In contrast, the Cauchy-fixing strategy fixes the Cauchy horizon at $\sigma \rightarrow \infty$. In both cases, we observed an accumulation of hyperboloidal hypersurfaces at the singular point, connecting $r = 0$, ${\cal C}_{\rm f}^+$, and $i^+$. This provides a geometrical explanation for the violation of constraints \eqref{eq:cond_Allternative_Heun} for the conformal Heun equation. We also showed that these arguments extend to any hyperboloidal foliation.

Despite the non-Heun form of the hyperboloidal radial equation, the spacetime framework retains a structure that can still be analysed using methods akin to those applied in the Heun function framework. In other words, the choice of a height function in the spacetime coordinate transformation automatically determines the mapping at the level of the radial equation. In the hyperboloidal framework, this mapping remains homotopic, as it continuously maps functions into functions. However, it is more general than s-homotopic transformations, which restrict the mapping to a subspace of the functional space. In particular, one can infer from the Heun function the resulting characteristic exponents \eqref{eq:CharExp_rho} and \eqref{eq:CharExp_sigeta} for the hyperboloidal radial solution. As expected, the coefficients describing the behaviour at the black hole horizon ${\cal H}_{\rm f}^+$ and future null infinity $\scri^+$ are trivial, since the time slices are naturally adapted to these regions. Furthermore, the framework provides the expected behaviour towards the black hole horizon ${\cal H}_{\rm p}^-$ and past null infinity $\scri^-$, laying the groundwork for a research programme that expands the hyperboloidal framework. In particular, it serves as a starting point for developing numerical infrastructure to compute ``in" and ``up” homogeneous solutions, linking numerics with formal mathematical approaches~\cite{Gajic:2019oem, Gajic:2019qdd, Gajic:2024xrn}, and advancing a hyperboloidal formulation for quasinormal modes orthogonality relations~\cite{Zimmerman:2014aha, London:2020uva, Green:2022htq, London:2023aeo, London:2023idh}.

\subsection*{Acknowledgements}
MM is supported by the European Union’s Horizon Europe 2024 research and innovation programme under the Marie Sklodowska-Curie grant agreement No. 101154525.
RPM acknowledges support from the Villum Investigator program supported by the VILLUM Foundation (grant no. VIL37766) and the DNRF Chair program (grant no. DNRF162) by the Danish National Research Foundation and the European Union’s Horizon 2020 research and innovation programme under the Marie Sklodowska-Curie grant agreement No. 101131233.%
\begin{appendices}

\end{appendices}

\bibliography{bibitems}
\end{document}